\documentclass{amsart}

\usepackage{enumerate}
\usepackage{color}

\usepackage{amssymb}
\usepackage{latexsym}
\usepackage{array}
\usepackage{graphics}
\usepackage{epstopdf}
\usepackage{epsf,graphicx,subfigure}
\usepackage{multirow}
\usepackage{lineno,hyperref}
\modulolinenumbers[5]
\usepackage{tikz}
\usetikzlibrary{arrows,positioning,shapes.geometric}

\usepackage[english]{babel} % English language/hyphenation
\usepackage{amsmath,amsfonts,amsthm,bm} % Math packages
\usepackage[linesnumbered,commentsnumbered,ruled]{algorithm2e}
\usepackage{tabularx}
\usepackage{booktabs}

\definecolor{dark_grey}{gray}{0.7}

\newcolumntype{v}[1]{>{\raggedright\hspace{0pt}}p{#1}}
\newcolumntype{V}[1]{>{\scriptsize\raggedright\hspace{0pt}}p{#1}}

\begin{document}

\title[Diffuse to fuse EEG spectra]{Diffuse to fuse EEG spectra -- intrinsic geometry of sleep dynamics for classification}

\author[G.-R. Liu]{Gi-Ren Liu}
\address{Department of Mathematics, National Chen-Kung University, Tainan, Taiwan}

\author[Y.-L. Lo]{Yu-Lun Lo}
\address{Department of Thoracic Medicine, Healthcare Center, Chang Gung Memorial Hospital, Chang Gung University, School of Medicine, New Taipei, Taiwan}

\author[J. Malik]{John Malik}
\address{Department of Mathematics, Duke University, Durham, NC, USA}

\author[Y.-C. Sheu]{Yuan-Chung Sheu}
\address{Department of Applied Mathematics, National Chiao Tung University, Hsinchu, Taiwan}

\author[H.-T. Wu]{Hau-Tieng Wu}
\address{Department of Mathematics and Department of Statistical Science, Duke University, Durham, NC, USA. Mathematics Division, National Center for Theoretical Sciences, Taipei, Taiwan}
\email{hauwu@math.duke.edu}

\maketitle

\begin{abstract}
We propose a novel algorithm for sleep dynamics visualization and automatic annotation by applying diffusion geometry based sensor fusion algorithm to fuse spectral information from two electroencephalograms (EEG). The diffusion geometry approach helps organize the nonlinear dynamical structure hidden in the EEG signal. The visualization is achieved by the nonlinear dimension reduction capability of the chosen diffusion geometry algorithms. For the automatic annotation purpose, the {support vector machine} is trained to predict the sleep stage. The prediction performance is validated on a publicly available benchmark database, Physionet Sleep-EDF [extended] SC$^*$ {(SC = Sleep Cassette)} and ST$^*$ {(ST = Sleep Telemetry)}, with the leave-one-subject-out cross validation. When we have a single EEG channel (Fpz-Cz), the overall accuracy, macro F1 and Cohen's kappa achieve $82.72\%$,$75.91\%$ and $76.1\%$ respectively in Sleep-EDF SC$^*$ and $78.63\%$, $73.58\%$ and $69.48\%$ in Sleep-EDF ST$^*$. This performance is compatible {with} the state-of-the-art results. When we have two EEG channels (Fpz-Cz and Pz-Oz), the overall accuracy, macro F1 and Cohen's kappa achieve $84.44\%$,$78.25\%$ and $78.36\%$ respectively in Sleep-EDF SC$^*$ and $79.05\%$, $74.73\%$ and $70.31\%$ in Sleep-EDF ST$^*$. The results suggest the potential of the proposed algorithm in practical applications.
\end{abstract}

\section{Introduction}
Sleep is a universal recurring dynamical and physiological activity in mammals. According to American Academy of Sleep Medicine (AASM) \cite{Iber2007,berry2012aasm} that generalizes Rechtschaffen and Kales (R\&K) criteria \cite{Rechtschaffen_Kales:1968}, the sleep dynamics can be divided into two broad stages, the rapid eye movement (REM) and the non-rapid eye movement (NREM), and the NREM stage is further divided into shallow sleep (stage N1 and N2) and deep sleep (stage N3).
Up to now, we have accumulated a lot of physiological knowledge about sleep dynamics \cite{Saper2013} and a lot of research is actively carried out to explore the unknowns \cite{Kanda2016}.
Despite those unknowns, it has been well known that a distortion of sleep dynamics could lead to catastrophic outcomes. For example, {sleep deprivation impacts decision making \cite{harrison2000impact}}, REM disturbance slows down the perceptual skill improvement \cite{Karni1994}, deprivation of slow wave sleep is associated with Alzheimer's disease \cite{Kang2009}, insufficient N2 sleep is associated with weaning failure \cite{RocheCampo2010}, several public disasters are caused by low sleep quality \cite{horne1999vehicle,Leger_Bayon_Laaban_Philip:2012}, etc.
Therefore, in addition to the interest stemming from physiological aspects, we have a lot of clinical applications from knowing the sleep dynamics.

The polysomnography (PSG) is the gold standard of evaluating the sleep dynamics \cite{collop2007clinical}. However, scoring the overnight sleep stage from the PSG outputs by human experts is time consuming and error-prone due to the huge signal loading \cite{Norman2000}. Due to the advance of the technology and computational power, in the past decades a lot of effort has been devoted to establish an artificial intelligence (AI) system for the automatic sleep stage annotation purpose. 
In addition to being able to accurately score sleep stages, an ideal AI system should also be able to ``self-learn'' or ``accumulate knowledge'' from the historical database. In other words, when the database with experts' annotations grows, an ideal AI system should be able to {use those new annotations}. 
It is worth mentioning that the challenge is actually ubiquitous -- for a new-arriving subject, {how may one utilize} the existing database with the expert annotations?

\subsection{Related Work}\label{sec:relatedwork}

There have been many proposed algorithms for the sake of automatic sleep stage scoring.
Among many papers, we shall distinguish two common cross validation (CV) schemes. According to whether the training data and the testing data come from different subjects, the literature is divided into two groups, {\em leave-one-subject-out} and {\em non-leave-one-subject-out} CV. When the validation set and training set are determined {\em on the subject level}, that is, the training set and the validation set contain different subjects, we call it the leave-one-subject-out CV (LOSOCV) scheme; otherwise we call it the non-LOSOCV scheme. The main challenge of the LOSOCV scheme comes from the inter-individual variability, but this scheme is close to the real-world scenario -- how to predict the sleep dynamics of a new arrival subject from a given annotated database. On the other hand, in the non-LOSOCV scheme, the training set and the testing set are dependent, and the performance might be over-estimated.
In this work, to better evaluate the proposed algorithm caused by the inter-individual variability issue, we focus on the LOSOCV scheme, and only summarize papers considering the LOSOCV scheme and single- or two- channel EEG signal.

In \cite{Gudmundsson2005}, time-domain and frequency-domain features are designed to predict awake, REM, light NREM stage (N1 and N2 are combined) and deep sleep stage by support vector machine. The performance is evaluated in a private database with 4 subjects. In \cite{MemarFaradji2018}, 104 features are extracted from the EEG signals and selected features are used to classify 4 sleep stages (N1 and N2 are combined) by {random forests}. The following papers classify 5 sleep stages like our work.
In \cite{2016autoencoder}, time-domain and frequency-domain features are used as the input layer of the stacked sparse autoencoder, which is a specific type of neural network {model}, with the sigmoid activation function.
The performance of the stacked sparse autoencoder was evaluated in the publicly available benchmark database, Sleep-EDF SC$^{*}$ \cite{Goldberger_Amaral_Glass_Hausdorff_Ivanov_Mark_Mietus_Moody_Peng_Stanley:2000}, and the overall accuracy was 78.9\%.
Instead of extracting features based on the domain knowledge, features in \cite{tsinalis2016automatic} are automatically learned by the convolutional neural networks (CNNs). The overall accuracies was 74.8\% for the same Sleep-EDF SC$^{*}$ database.
In \cite{DeepSleepNet}, the authors also proposed a deep learning model, named DeepSleepNet, which utilized CNNs to extract features and apply the bidirectional Long Short-Term Memory to learn the stage transition rules from EEG epochs. The DeepSleepNet algorithm reaches the state-of-the-art 82.0\% of overall accuracy on the Sleep-EDF SC$^{*}$ database. In \cite{Vilamala2017}, a similar approach based on the deep CNN with modifications is considered and achieves a compatible result.

\subsection{Our contribution}

The main novelty and contribution of this paper is the {\em diffusion geometry} based {feature extraction when there is only one electroencephalogram (EEG) channel, and {\em diffusion geometry} based {\em sensor fusion} when there are two EEG channels. The goal is to extract intrinsic information hidden in the EEG.}   
This novel diffusion geometry approach achieves the above-mentioned challenges -- an accurate automatic sleep stage scoring AI system. %

{The proposed algorithm is composed of feature extraction part and learning part.} The feature extraction part of the algorithm is based on a combination of two modern signal processing tools, including the {scattering transform \cite{mallat2012group}} and diffusion map (DM) \cite{coifman2006diffusion} {when only single channel is available} or diffusion-based sensor fusion (the multiview DM \cite{Linderbaum_Yeredor_Salhov_Averbuch:2015}) {when two channels are available}. 
The {scattering transform is} a nonlinear-type {signal processing tool motivated by studying the convolutional neural network (CNN) with solid theoretical supports, which allows us an extraction of useful features from the EEG signal. We mention that while it is motivated by studying CNN, the label is not used in the scattering transform, so it is an unsupervised learning tool}. We call the {extracted features {\em scattering} EEG spectral features}.
The {scattering} EEG spectral feature, however, is in general {of high dimension and} different from the intrinsic sleep dynamics. This difference comes from various resources, including, for example, the inevitable noise and artifact. Thus, an extra step is needed to {reduce the dimension} and refine/reorganize the scattering EEG spectral feature, which leads to the final intrinsic features for the sleep dynamics. We suggest DM \cite{coifman2006diffusion} or diffusion-based sensor fusion algorithms, like multiview diffusion \cite{Linderbaum_Yeredor_Salhov_Averbuch:2015}.  
In short, the scattering EEG spectral features extracted by the scattering transform are re-organized by modern diffusion geometry based sensor fusion algorithms. We call the output features of the DM/diffusion-based sensor fusion the {\em intrinsic sleep dynamical features}.
The learning part of the algorithm for the prediction purpose is based on the {kernel support vector machine (SVM) \cite{Scholkopf_Smola:2002}}.
{The prediction model for each testing subject is established by the kernel SVM based on the intrinsic sleep dynamical features and the annotations of all remaining subjects in the database.
}

To show this performance of the proposed algorithm and have a fair comparison with reported results, we consider the publicly available benchmark database, the PhysioNet Sleep-EDF SC$^{*}$ {(SC = Sleep Cassette)} database \cite{Goldberger_Amaral_Glass_Hausdorff_Ivanov_Mark_Mietus_Moody_Peng_Stanley:2000}, which consists of healthy subjects without any sleep-related medication, and the PhysioNet Sleep-EDF ST$^{*}$ {(ST = Sleep Telemetry)} \cite{Goldberger_Amaral_Glass_Hausdorff_Ivanov_Mark_Mietus_Moody_Peng_Stanley:2000}, which consists of subjects who had mild difficulty falling asleep. {We mention that compared with the existing neural network (NN) based algorithms mentioned in Section \ref{sec:relatedwork}, the proposed algorithm, including the feature extraction part and learning part, has solid theoretical backups. In addition, later in this paper we will show that the proposed algorithm performs better or at least equally to the existing reported results based on NN approaches when there is a single EEG channel, and the performance is improved if we fuse information from two EEG channels.}

\subsection{Organization of the paper}
The rest of this paper is organized as follows. In Section \ref{sec:feature_extraction}, we introduce the proposed algorithm and the databases we evaluate the proposed algorithm. In Section \ref{sec:feature_extraction_Scattering} we summarize the theoretical background of the scattering transform and introduce the {\em scattering EEG spectrum} as the dynamical features. In Section \ref{sec:DM}, we describe how to apply the diffusion geometry to integrate and fuse the dynamical features in Section \ref{sec:feature_extraction_Scattering}, and the prediction model {based on SVM} is summarized in Section \ref{sec:HMM}.
In Section \ref{sec:description_database}, we describe two databases, including the Sleep-EDF Database [Expanded] from PhysioNet. 
The results of applying the proposed algorithm to the two databases are shown in Section \ref{sec:experiments}.

\section{Material and Method}\label{sec:feature_extraction}

The main novelty of the proposed algorithm is the feature extraction that depends on the diffusion geometry based sensor fusion tools, DM and multiview DM. 
The feature extraction consists of two steps. First, we apply {the scattering transform to} extract features from the EEG signal (indicated by Part 1 in Figure \ref{FlowChart}). Second, we apply DM or multiview diffusion map (indicated by Part 2-1 and Part 2-2 in Figure \ref{FlowChart}) with a properly designed metric to determine the final features. 
With the proposed features, we take the well established {SVM} to build up the prediction model. 
Below, we detail the algorithm implementation step by step.

\tikzstyle{line} = [draw, -latex']
\tikzstyle{arrow} = [thick,->,>=stealth]

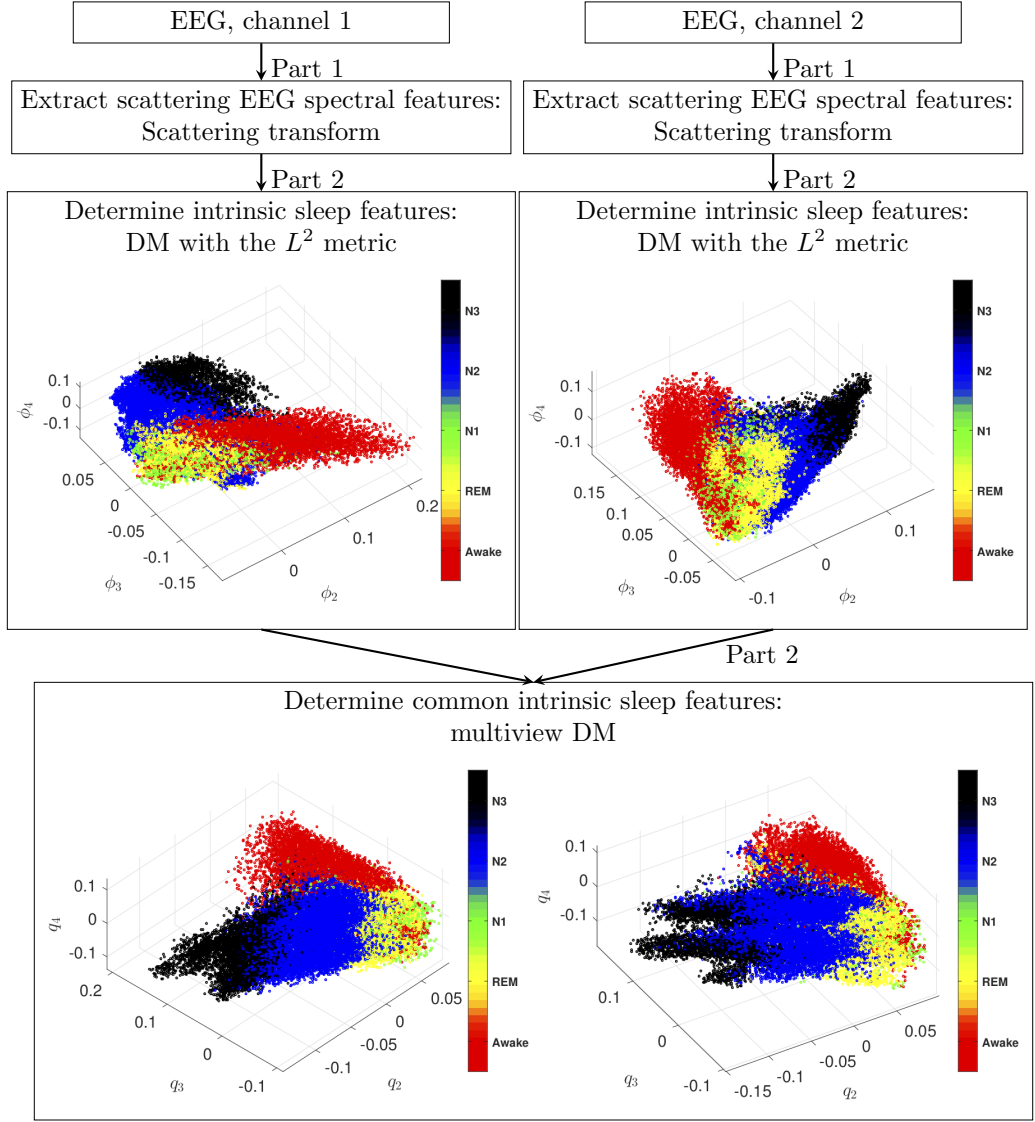
\begin{figure}[h!]
\centering
   \begin{tikzpicture}[>=latex']
        \tikzset{
        block/.style= {draw, rectangle, align=center,minimum width=5cm,minimum height=.10cm,line width=0.05mm},
        rblock/.style={draw, shape=rectangle,rounded corners=1.5em,align=center,minimum width=12cm,minimum height=.10cm},
        }

        \node [block]  (EEG1start) {EEG, channel 1};
        \node [block, below =.5cm of EEG1start] (EEG1Step1) {Extract scattering EEG spectral features: \\
        Scattering transform};
        \node [block, below =.5cm of EEG1Step1] (EEG1Step2) {Determine intrinsic sleep features: \\ DM with the $L^{2}$ metric \\ \includegraphics[scale=0.44]{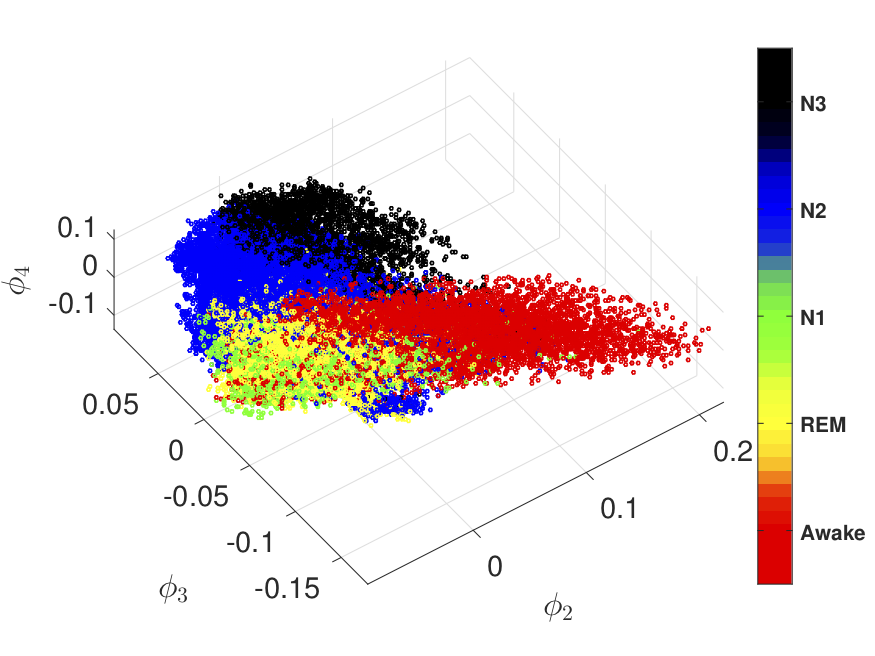}};

        \node [block, right =1.8cm of EEG1start]  (EEG2start) {EEG, channel 2};
        \node [block, below =.5cm of EEG2start] (EEG2Step1) {Extract scattering EEG spectral features: \\
        Scattering transform};
        \node [block, below =.5cm of EEG2Step1] (EEG2Step2) {Determine intrinsic sleep features: \\ DM with the $L^{2}$ metric \\\includegraphics[scale=0.44]{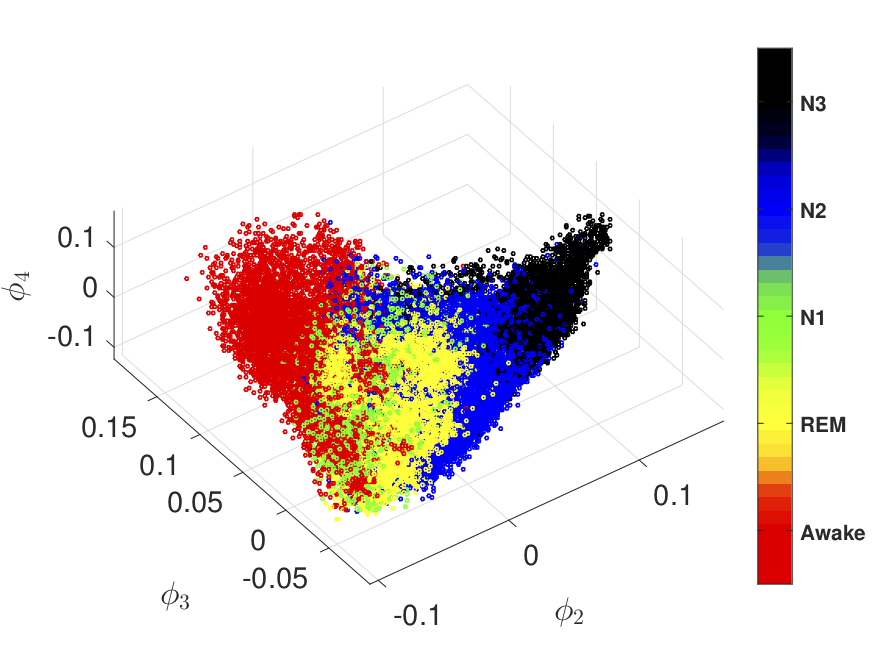}};

        \node [block, below right =.7cm and -6.4 cm of EEG1Step2] (Final) {Determine common intrinsic sleep features: \\
        multiview DM \\
        \includegraphics[scale=0.44]{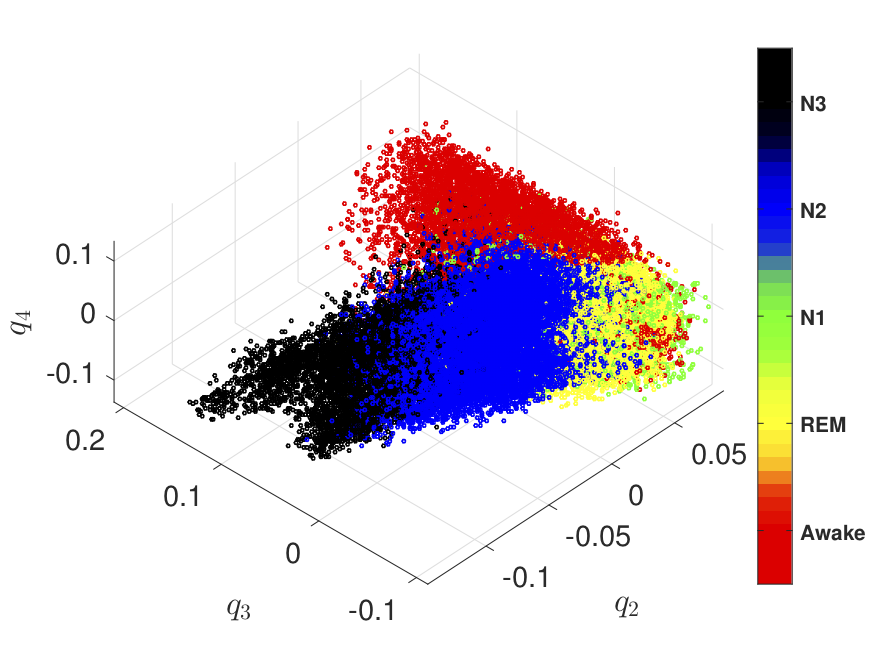}\includegraphics[scale=0.44]{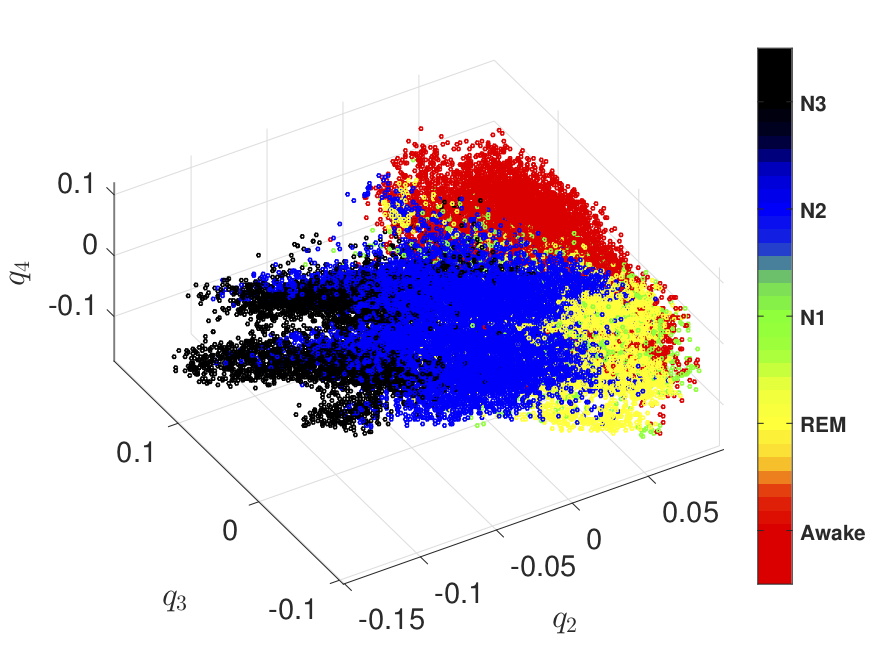}};

          \draw[arrow]   (EEG1start.south)  --(EEG1Step1.north) node [pos=0.66,right] {Part 1};
          \draw[arrow]   (EEG1Step1.south)  --(EEG1Step2.north) node [pos=0.66,right] {Part 2};
          \draw[arrow]   (EEG1Step2.south)  -- (Final.north) node [pos=0.66,right] {};

                    \draw[arrow]   (EEG2start.south)  --(EEG2Step1.north) node [pos=0.66,right] {Part 1};
          \draw[arrow]   (EEG2Step1.south)  --(EEG2Step2.north) node [pos=0.66,right] {Part 2};
          \draw[arrow]   (EEG2Step2.south)  -- (Final.north) node [pos=0.46,right] {\quad\quad Part 2};

    \end{tikzpicture}
    \caption{\label{FlowChart}The flow chart of the proposed feature extraction steps and a visualization of the extracted features by the diffusion map and sensor fusion are shown. The signal is from 20 subjects in Sleep-EDF SC$^{*}$ \cite{Goldberger_Amaral_Glass_Hausdorff_Ivanov_Mark_Mietus_Moody_Peng_Stanley:2000}, where Channel 1 is Fpz-Cz and channel 2 is Pz-Oz. In the bottom figure, only one of the multiview DMs is shown. The ratios of the stages Awake, REM, N1, N2, and N3 are 19\%, 18\%, 7\%, 42\%, and 14\%.}
\end{figure}

\subsection{Feature extraction step, part 1: Scattering transform}\label{sec:feature_extraction_Scattering}

In this paper, we apply the {recently developed scattering transform \cite{mallat2012group} to extract spectral features  \cite{anden2014deep} from a given EEG signal.
The scattering transform is motivated by establishing a mathematical foundation of the convolutional neural network. It computes coefficients of the given EEG signal by iteratively applying wavelet transforms and nonlinear modulus operators.
They have two benefits. First,
they are invariant to time-shifts and stable to time-warping deformation \cite{anden2014deep}.
Second, they can effectively characterize self-similarity and intermittency properties of multiscale time series \cite{bruna2015intermittent}.
These two benefits provide us with a better quantification of the EEG dynamics.
Here, we briefly review the scattering transform, particularly the first- and second-order coefficients, which we use as the spectral features for the given EEG signal.

Take a signal $x$ defined on $[0,T]$. 
Let $\psi$ be an analytic wavelet, that is, a complex band-pass filter with the support over positive frequencies and the central frequency $1$. %
Denote the associated father wavelet by $\phi$, which is a low pass filter.
Denote $Q\in \mathbb{N}$ to be the number of wavelets per octave; that is, the bandwidth of $\hat{\psi}$ is of order $Q^{-1}$. 
To evaluate the wavelet coefficients, we dilate the mother wavelet $\psi$ as follows:
\begin{equation}
\psi_{j}(t) = 2^{-j/Q}\psi(2^{-j/Q}t)\,,
\end{equation}
where $j\in\{\ldots,-1,0,1,\ldots, H\}$, and $H\in \mathbb{N}$ is the maximal wavelet scale so that the largest-scale wavelet will be of bandwidth $T$. Clearly, $\widehat{\psi_j}$ is centered at $2^{-j/Q}$ with the bandwidth $2^{-j/Q}Q^{-1}$, and hence $\psi_{H}$ captures the frequency above $2^{-H/Q}$. For the low frequency region $[0,2^{-H/Q}]$, we follow the suggestion in \cite{anden2014deep} to cover this region by $Q-1$ equally spaced filters, denoted as $\psi_{H+1},\ldots,\psi_{H+Q}$, with the frequency bandwidth $2^{-H/Q}Q^{-1}$ centered at $l2^{-H/Q}Q^{-1}$, where $l=1,\ldots,Q-1$. Again, we follow the suggestion in \cite{anden2014deep} and call $\psi_{H+1},\ldots,\psi_{H+Q-1}$ wavelets. Denote $\Lambda=\{2^{-j/Q}\}_{j=-\infty}^H\cup\{l2^{-H/Q}Q^{-1}\}_{l=1}^{Q-1}$ to be the grid of center frequencies of all considered wavelets. 

With the above preparation, we are ready to establish the scattering transform coefficients of the signal $x$. We start from collecting ordinary wavelet transform coefficients $x\star\psi_{j_1}(t)$, where $\star$ is the convolution operator and $-\infty\leq j_1\leq H+Q$, and the moving average by the father wavelet $x\star\phi_H(t)$, where 
\begin{equation}
\phi_H(t):=Q^{-1}2^{-H/Q}\phi(Q^{-1}2^{-H/Q}t)\,. 
\end{equation}
Note that $\widehat{\phi_H}$ has the frequency bandwidth $2^{-H/Q}Q^{-1}$. 
Denote 
\begin{equation}
S_0x(t):=x\star\phi_H(t)\,,
\end{equation}
which is called the {\em zeroth-order coefficients of the scattering transform}. Clearly, $S_0x(t)$ contains only the low frequency information of the signal $x$.
On the other hand, while the wavelet transform, $x\star \psi_j(t)$, provides multi-scale spectral information of $x$ at the scale $2^{-j/Q}$,
the scattering transform outputs translation-invariant coefficients by averaging the modulus of the complex coefficients $x\star \psi_j(t)$ via the following moving average step:
\begin{equation}\label{Definition:1st ST}
{S}_{1}x(t,j_1):=|x\star\psi_{j_1}|\star\phi_H\,,
\end{equation}
where 
$-\infty< j_1\leq H$. We call $S_1x(t,j_1)$ the {\em first-order coefficients of the scattering transform}. 

The time averaging operator in \eqref{Definition:1st ST} results in the loss of the fine-scale information in $|x\star\psi_{j_{1}}|$ such as vibratos and attacks \cite{anden2014deep}.
To retrieve the lost information, we compute $|x\star\psi_{j_{1}}|\star\psi_{j_{2}}$. Again, to outputs translation-invariant coefficients, we apply modulation and convolution operators on $|x\star\psi_{j}|\star\phi_H$ via:
\begin{equation}
{S}_{2}x(t,j_{1},j_{2}):=||x\star\psi_{j_{1}}|\star\psi_{j_{2}}|\star\phi_H\,,
\end{equation}
where $-\infty\leq j_1<j_2\leq H$.
Note that only the wavelet $\psi_{j_{2}}$ with larger scale $j_{2}>j_{1}$ are needed, since $|x\star\psi_{j_{1}}|$ has approximately the same frequency support as that of $\psi_{j_{1}}$ \cite[(29)]{anden2014deep}.
The outputs are called the {\em second-order coefficients of the scattering transform} of $x$. This iterative procedure can be extended to higher orders. For example, the third-order  coefficients of the scattering transform of $x$ can be defined by evaluating
$|||x\star\psi_{j_{1}}|\star\psi_{j_{2}}|\star \phi_{j_3}|\star\phi_H$, 
where $-\infty\leq j_1<j_2<j_3\leq H$. In this paper, only the first-order and second-order coefficients are exploited.

Next, we follow the suggestions in \cite{anden2014deep} to post-process the scattering transform coefficients. To de-correlate the correlations at different orders \cite{anden2014deep}, the first-order scattering coefficients are renormalized as follows:
\begin{equation}
\widetilde{S}_{1}x(t,j_{1}) = \frac{|x\star\psi_{j_{1}}|\star\phi_H(t)}{|x              |\star\phi_H(t)+\varepsilon},
\end{equation}
where $\varepsilon = 2^{-20}$, and
the 2nd-order scattering coefficients are renormalized by coefficients of the 1st-order:
\begin{equation}
\widetilde{S}_{2}x(t,j_{1},j_{2}) = \frac{||x\star\psi_{j_{1}}|\star\psi_{j_{2}}|\star\phi_J(t)}{|x\star\psi_{j_{1}}|\star\phi_H(t)+\varepsilon},
\end{equation}
where $\varepsilon = 2^{-20}$. 
Similar to the Mel-frequency cepstral coefficients, we follow \cite{anden2014deep} and apply a logarithm to $x\star\phi_H(t)$ and the normalized coefficients, i.e., $\log\widetilde{S}_{1}x(t,j_{1})$ and $\log\widetilde{S}_{2}x(t,j_{1},j_{2})$.
According to \cite{anden2014deep}, this step can transfer the multiplicative components into additive ones and improve the classification performance.

We now apply the scattering transform to extract features from the EEG signal. Following the R\&K \cite{Rechtschaffen_Kales:1968} and AASM classifications \cite{berry2012aasm},
we take an epoch for the sleep stage evaluation to be 30 seconds long.
Suppose the $l$-th epoch starts at a certain time $t_{l}$ and ends at $t_{l}+30$.
Let $x_l(t)$ be a 90s EEG signal segment during the period $[t_{l}-60,t_{l}+30]$; that is, we consider 60s extra EEG signal before the epoch we have interest in and hence the signal length is $T=90$s. The EEG signal is sampled uniformly every $\tau$ second, where $\tau$ is the reciprocal of the sampling rate. For example, $\tau=1/100$ for the Sleep-EDF Database.
In this work, we consider the wavelet $\psi$ to be the Morlet mother wavelet, and take $Q=2$. 
Numerically, we represent $x_k$ as a $9,000$-dim vector and take $H=17$, $0\leq j_1\leq H$ for $S_1x_k$, and $0\leq j_1<j_2\leq H$ for $S_2x_k$. 
Since $S_0x_l$, $S_1x_l$ and $S_2x_l$ are redundant at the time axis, they are subsampled at the rate $2^{H/Q-1}$; that is, over half-overlapping time windows of size $2^{H/Q}$ centered at the points $t_l:=l2^{H/Q-1}$, where $l\in \mathbb{N}$. 
The collection of these subsampled coefficients are concatenated and called the {\em scattering EEG spectral feature} of the $k$-th epoch, which we denote as $\mathbf{u}^{(k)}$.
}

\subsection{Feature extraction step, part 2: (Multiview) diffusion map}\label{sec:DM}

In the optimal situation, the sleep dynamics at each epoch can be well captured by the {scattering EEG spectral features}. However, the {scattering EEG spectral features} might be erroneous due to the inevitable noise, other sensor-specific artifacts and the information distortion caused by the observation procedure. We then stabilize these features to better quantify the intrinsic sleep dynamics
by applying DM. DM is a nonlinear manifold learning algorithm that not only is robust to noise but also respects the nonlinear structure of the intrinsic dynamics.
We detail its implementation below, together with its generalization, multiview DM, to fuse information from multiple channels.

\subsubsection{Extract intrinsic sleep dynamics by diffusion map}

Take the {scattering EEG spectral features $\mathcal{U}^{x,k}:=\{\mathbf{u}^{(j)}\}_{j=1}^{J_k}$} generated from the single-channel EEG signal $x(t)$ {of the $k$-th subject, where $k=1,\ldots,K$, which is a point cloud in an Euclidean space. Denote $\mathcal{U}^x:=\cup_{k=1}^K\mathcal{U}^{x,k}=\{\mathbf{u}^{(j)}\}_{j=1}^{J}$, where $J=\sum_{k=1}^KJ_k$ and $\{\mathbf{u}^{(J_1+\ldots+J_{k-1}+1)},\ldots,\mathbf{u}^{(J_1+\ldots+J_{k})}\}$ comes from $\mathcal{U}^{x,k}$.} 
First, from the point cloud $\mathcal{U}^x$, we build a graph with $\mathcal{U}^x$ being vertices. The affinity between features $\mathbf{u}^{(i)}$ and $\mathbf{u}^{(j)}$ is defined as
\begin{equation}\label{affinity_matrix}
W_x(i,j)=\exp\left\{-\frac{\|\mathbf{u}^{(i)}-\mathbf{u}^{(j)}\|}{\varepsilon}\right\},\ \ \textup{for}\ \ i,j=1,\ldots,J\,,
\end{equation}
where $\epsilon>0$ is chosen by the user and {$\|\cdot\|$ is the Euclidean distance comparing two scattering EEG spectral features}.
Thus, we have a $J\times J$ affinity matrix $W_x$. Next, define the degree matrix $D_x$ of size $J\times J$, which is diagonal, as
\begin{equation}\label{degree_matrix}
D_x(i,i)=\sum_{j=1}^JW_x(i,j),\ \ \textup{for}\ \ i=1,\ldots,J.
\end{equation}
With matrices $W_x$ and $D_x$, we could define a random walk on the point cloud $\mathcal{U}^x$ with the transition matrix given by the formula
\begin{align}\label{Definition:Atransition}
A_{x}:=D_x^{-1}W_x\,.
\end{align}
Since $A_{x}$ is similar to the symmetric matrix $D_x^{-1/2}W_xD_x^{-1/2}$, it has a complete set of right eigenvectors $\phi_1, \phi_2,\cdots,\phi_{J}\in \mathbb{R}^J$ with corresponding eigenvalues $1 = \lambda_1 > \lambda_2\geq\cdots\geq\lambda_{J}\geq 0$, where $\phi_1 = [1, 1, \ldots , 1]^{\textup{T}}\in \mathbb{R}^J$. Indeed, from the eigen-decomposition $D_x^{-1/2}W_xD_x^{-1/2}=O\Lambda O^\top$, where $O$ is a $J\times J$ orthonormal matrix and $\Lambda$ is a $J\times J$ diagonal matrix, we have $A_x=U\Lambda V^\top$, where $U=D_x^{-1/2}O$ and $V=D_x^{1/2}O$.
With the decomposition $A_x=U\Lambda V^\top$, the DM is defined as
\begin{equation}\label{DM}
\Phi^x_t:\mathbf{u}^{(j)} \mapsto \big(\lambda_2^t\phi_2(j), \lambda_3^t\phi_3(j),\ldots,\lambda_{\hat{d}+1}^t\phi_{\hat{d}+1}(j)\big)\,,
\end{equation}
where $j=1,\ldots,J$ and $t>0$ is the diffusion time chosen by the user, and $\hat{d}$ is an estimate of the dimension of the intrinsic state. Here, $\hat{d}$ can be obtained according to the spectral gap in the decay of the eigenvalues $\{\lambda_{j}\}_{j=1}^{J}$. In other words, $\Phi^x_t(\mathbf{u}^{(j)})$ consists of the second to $\hat{d}+1$ coordinates of $e_j^\top U\Lambda^t$, where $e_j$ is the unit $J$-dim vector with the $j$-th entry $1$.
As a result, the {scattering EEG spectral features} $\mathcal{U}^x$ are converted into a set of new features in $\mathbb{R}^{\hat{d}}$ via the DM. Denote $\mathcal{F}^x:=\{\Phi^x_t(\mathbf{u}^{(j)}) \}_{j=1}^J\subset \mathbb{R}^{\hat{d}}$. {Therefore, we obtain the {\em intrinsic sleep features} for the recorded EEG signal $x$ of the $k$-th subject $\mathcal{F}^{x,k}:=\{\Phi^x_t(\mathbf{u}^{(J_1+\ldots+J_{k-1}+1)}),\ldots,\Phi^x_t(\mathbf{u}^{(J_1+\ldots+J_{k})})\}$.}

Note that the diffusion process plays an essential role in the whole algorithm, since (\ref{DM}) is based on the diffusion nature of the random walk with the transition matrix (\ref{Definition:Atransition}). Based on the spectral geometry result reported in \cite{Berard_Besson_Gallot:1994}, and the asymptotical spectral convergence results shown in \cite{VonLuxburg_Belkin_Bousquet:2008,Singer_Wu:2016}, we know that the Riemannian manifold hosting the inaccessible intrinsic sleep features is almost isometrically recovered by DM (\ref{DM}). 
In addition to recovering the intrinsic state space, another important property of DM is its robustness to noise. Essentially, the diffusion process could be viewed as a nonlinear ``averaging'' process. This viewpoint was first captured and proved in \cite{ElKaroui:2010a}, and later extended to larger noise in \cite{ElKaroui_Wu:2016b}. {Another important fact is that the embedding provided by DM is unique up to a global rotation, so it is not feasible to construct intrinsic sleep features for each subject separately and pool them together for the prediction purpose. The proposed algorithm here applies DM to all scattering EEG spectral features from all subjects to construct intrinsic sleep features, which is over a {\em universal coordinate} that allows us a comparison among different subjects.}
See Figure \ref{FlowChart} for an example of DM when $\hat{d}=3$. It is clear that epochs of different sleep stages are clustered and well separated.

\subsubsection{Fuse two channels via multiview DM}

If only one EEG signal $x$ is available, we proceed to the learning step in Section \ref{sec:HMM} with the intrinsic sleep features $\mathcal{F}^x$ for channel $x$. If we have two or more channels, we can take them into account simultaneously.
In general, this problem is understood as the {\em sensor fusion} problem. While different EEG channels capture information from the same brain, the information recorded might vary and they might be contaminated by brain-activity irrelevant artifacts from different {sources}, including noise and other sensor-specific nuisance. These artifacts not only deteriorate the quality of the extracted features but might also mislead the analysis result. The main purpose of sensor fusion is distilling the brain information and removing those unwanted artifacts. A naive way to combine information from $x$ and $y$ is simply concatenating features and form a new feature. However, due to the inevitable sensor-specific artifacts or errors, such a concatenating scheme might not be the optimal route to fuse sensors \cite{lederman2015alternating,Talmon_Wu:2016}.
We consider the recently developed diffusion-based sensor fusion approach, multiview DM \cite{Linderbaum_Yeredor_Salhov_Averbuch:2015} or equivalently the alternating DM \cite{Talmon_Wu:2016}, to fuse sensor information. 
Essentially, the information from different sensors are ``diffused'' to integrate the common information, and simultaneously eliminate artifacts or noise specific to each sensor.

{In this work, we focus on} two simultaneously recorded EEG channels $x$ and $y$. Here we detail the multiview DM algorithm. For each channel, we apply Sections \ref{sec:feature_extraction_Scattering} and \ref{sec:DM} to obtain its intrinsic sleep features, denoted as $\mathcal{F}^x \subset \mathbb{R}^{\hat{d}_x}$ and $\mathcal{F}^y \subset \mathbb{R}^{\hat{d}_y}$ respectively, where $\hat{d}_x$ might be different from $\hat{d}_y$.
Define
\begin{equation}\label{equ:multiview2}
W_{x,y} =
\begin{bmatrix}
0_{J\times J} & W_xW_y\\
W_yW_x & 0_{J\times J}
\end{bmatrix}\in \mathbb{R}^{2J\times 2J},
\end{equation}
where $W_x$ and $W_y$ are affinity matrices defined in \eqref{affinity_matrix}, and define a diagonal matrix $D_{x,y} \in \mathbb{R}^{2J\times 2J}$ so that its $i$-th diagonal entry is the sum of the $i$-th row of $W_{x,y}$. Denote $q_i\in \mathbb{R}^{2J}$ to be the $i$-th right eigenvector of the transition matrix $D_{x,y}^{-1}W_{x,y}$
corresponding to eigenvalue $\sigma_{i}$. 
Since for each $i$, $q_i(l)$ and $q_i(J+l)$ correspond to the $l$-th epoch for each $l\in\{1,\ldots,J\}$,
we call the $2\tilde{d}$-dimensional vector 
\begin{equation}\label{Definition:common intrinsic sleep feature}
v_j:=[\sigma_{2}^{t}q_2(j)\ \cdots\ \sigma_{\tilde{d}+1}^{t}q_{\tilde{d}+1}(j)\ \sigma_{2}^{t}q_2(J+j)\ \cdots\ \sigma_{\tilde{d}+1}^{t}q_{\tilde{d}+1}(J+j)]^\top
\end{equation} 
the {\em common intrinsic sleep feature} associated with the sleep stage of the $j$-th epoch, where $j\in \{1,\ldots,J\}$. 
Denote $\mathcal{F}^{x,y}:=\{v_j\}_{j=1}^J$. {Hence, we obtain the common intrinsic sleep features for the simultaneously recorded EEG signals $x$ and $y$ of the $k$-th subject $\mathcal{F}^{x,y,k}:=\{v^{(J_1+\ldots+J_{k-1}+1)},\ldots,v^{(J_1+\ldots+J_{k})}\}$.} 

{Note that since the transition matrix $D_{x,y}^{-1}W_{x,y}$ is associated with the random walk on the bipartite graph formed from epochs of $x$ and $y$, its eigenvectors ``co-clustering'' two channels \cite{dhillon2001co}. Hence, we can use its eigenvectors as features for the sleep stage. We mention that when there are only two channels, like in our case, the multiview DM is equivalent to alternating DM \cite{Talmon_Wu:2016}, but when there are more than two channels, the multiview DM is different from the generalization of alternating DM to multiple channels \cite{Katz_Wu_Lo_Talmon:2016}.}
An illustration of the result of multiview DM with $\tilde{d}=3$ is shown at the bottom of Figure \ref{FlowChart}.

{\subsection{Learning step: Sleep Stage Classification by the support vector machine}\label{sec:HMM}

In this work, we choose the standard and widely used kernel SVM \cite{Scholkopf_Smola:2002} for the learning step. SVM find a hyperplane in the feature space that separates the data set into two disjoint subsets. Based on the reproducing kernel Hilbert space theory, SVM is generalized to the {\it kernel SVM}, which allows for classification when there exists a nonlinear structure in the feature space. Specifically, kernel SVM finds a nonlinear surface separating the data set into two disjoints subsets. We refer the interested reader to \cite{Scholkopf_Smola:2002} for technical details.  

In this work, we choose the radial based function, $K(x,x')=\exp(-\frac{\|x-x'\|^2_2}{2\sigma^2})$, where $\sigma>0$, as the kernel function. In this sleep dynamics classification problem, the label is multi-class so we need to further generalize the kernel SVM to the multi-class SVM to complete our mission. To this end, we apply the one-versus-all (OVA) classification scheme \cite{Rifkin_Klautau:2004}. }

\subsection{Material}\label{sec:description_database}

To evaluate the proposed algorithm, we consider the commonly considered benchmark database, Sleep-EDF Database [Expanded], from the public repository Physionet \cite{Goldberger_Amaral_Glass_Hausdorff_Ivanov_Mark_Mietus_Moody_Peng_Stanley:2000}. It contains two subsets (marked as SC$^{*}$ and ST$^{*}$). The first subset SC$^{*}$ comes from healthy subjects without any sleep-related medication. The subset SC$^{*}$ contains Fpz-Cz/Pz-Oz EEG signals recorded from 10 males and 10 females without any sleep-related medication, and the age range is 25-34 year-old. There are two approximately 20-hour recordings per subject, apart from a single subject for whom there is only a single recording.
The EEG signals were recorded during two subsequent day-night periods at the subjects' home. The sampling rate is 100 Hz.
The second subset ST$^{*}$ was obtained in a 1994 study of temazepam effects on the sleep of subjects with mild difficulty falling asleep. The subset ST$^{*}$ contains Fpz-Cz/Pz-Oz EEG signals recorded from 7 males and 15 females, who had mild difficulty falling asleep. Since this data set is originally used for studying the effects of temazepam, the EEG signals were recorded in the hospital for two nights, one of which was after temazepam intake. Only their placebo nights can be downloaded from \cite{Goldberger_Amaral_Glass_Hausdorff_Ivanov_Mark_Mietus_Moody_Peng_Stanley:2000}. The sampling rate is 100 Hz.
For both SC$^*$ and ST$^*$ sets, each 30s epoch of EEG data has been annotated into the classes Awake, REM, N1,N2, N3 and N4. The epochs corresponding to movement and unknown stages were excluded and the the epochs labeled by N4 are relabeled to N3 according to the AASM standard.
For more details of the database, we refer the reader to \url{https://www.physionet.org/physiobank/database/sleep-edfx/}.

\subsection{Statistics}

To evaluate the performance of the proposed automatic sleep scoring algorithm, we consider the LOSOCV scheme. For each database, one subject is randomly chosen as the testing set and the remaining subjects form the training set. {Specifically, when there are $n$ subjects, for the $\ell$-th subject, we train the kernel SVM model using the common intrinsic sleep features $\cup_{k=1,\ldots,K,\,k\neq \ell}\mathcal{F}^{x,y,k}$ with the provided labels, and test the trained model on $\mathcal{F}^{x,y,\ell}$ and report the results. Then apply the same procedure for subject $\ell=1,\ldots,n$.} Note that this LOSOCV scheme helps prevent overfitting in constructing the prediction model.

All performance measurements used in this paper are computed through
the unnormalized confusion matrix ${ M}\in \mathbb{R}^{5\times 5}$.
For $1\leq p,q\leq 5$, the entry $M_{pq}$ represents the number of expert-assigned $p$-class epochs, which were predicted to the $q$-class.
The precision ($\textup{PR}_p$), recall ($\textup{RE}_p$), and F1-score ($\textup{F1}_p$) of the $p$-th class, where $p=1,\ldots,5$, are computed respectively through
\begin{equation}
\textup{PR}_{p}= \frac{M_{pp}}{\sum_{q=1}^5M_{qp}},\quad
\textup{RE}_{p}= \frac{M_{pp}}{\sum_{q=1}^5M_{pq}},\quad
\textup{F1}_{p}=\frac{2 \textup{PR}_{p}\cdot\textup{RE}_{p}}{\textup{PR}_{p}+\textup{RE}_{p}}\,.
\end{equation}
The overall accuracy (ACC), macro F1 score ($\textup{Macro-F1}$) and kappa ($\kappa$) coefficient are computed respectively through
\begin{equation}
\textup{ACC}= \frac{\sum_{p=1}^5  M_{pp}}{\sum_{p,q=1}^5M_{pq}   },\quad
\textup{Macro-F1}=\frac{1}{5}\sum_{p=1}^5\textup{F1}_{p},\quad
\kappa = \frac{\textup{ACC}-\textup{EA}}{1-\textup{EA}},
\end{equation}
where {EA means the expected accuracy,} which is defined by
\begin{equation}
\textup{EA}= \frac{\sum_{p=1}^5\left(   \sum_{q=1}^5 M_{pq}\right)\times \left(\sum_{q=1}^5M_{qp}\right)}{\left(\sum_{p,q=1}^5     M_{pq}\right)^{2}}.
\end{equation}

To evaluate if two matched samples have the same mean, we apply the one-tail Wilcoxon signed-rank test under the null hypothesis that the difference between the pairs follows a symmetric distribution around zero. 
When we compare the variance, we apply the one-tail F test under the null hypothesis that there is no difference between the variances. 
We consider the significance level of $0.05$. To handle the multiple comparison issue, we consider the Bonferroni correction.

\section{Results}\label{sec:experiments}

{In this section, we report the results of applying the proposed algorithm to the above-mentioned two databases. We show a visualization of the (common) intrinsic sleep features. Note that this can be viewed as a dimension reduction of the EEG signals. The dynamics of the intrinsic sleep features are also shown to compared with the hypnogram. The confusion matrices of the automatic sleep stage classification on different databases are shown, and a performance comparison with other reported results is also provided.}
The parameters in the numerical implementation are listed here. For DM and multiview DM, the $\epsilon$ in (\ref{affinity_matrix}) is chosen to be the $1$\% percentile of pairwise distances, the diffusion time is $t=0.3$.  $\hat{d}$ and $\tilde{d}$ are chosen to be 80. No systematic parameter optimization is performed. For the reproducibility purpose, the Matlab code will be provided via request.

\subsection{Sleep-EDF database (SC*)}

\begin{figure}[!htb]
\centering
\includegraphics[width=1\textwidth]{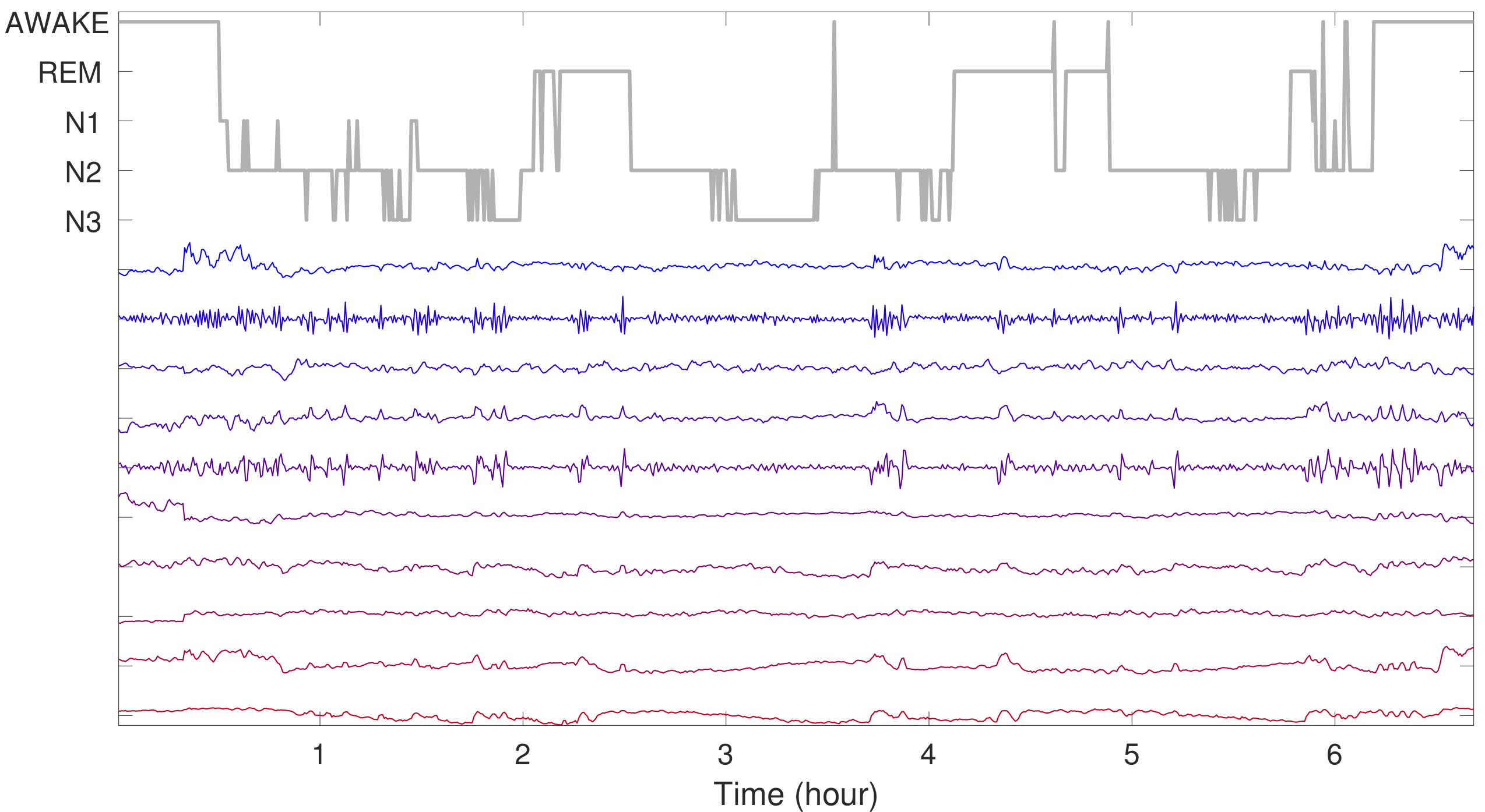}
\caption{A visualization of the temporal dynamics in common intrinsic sleep features extracted by the multiview DM. The first intrinsic feature is colored in
red, and the tenth intrinsic is colored in blue. The expert's labels are plotted on the top.}
\label{figure4:intrinsic_feature_dynamic}
\end{figure}

We start from showing the visualization of the intrinsic sleep features and the common intrinsic sleep features from the total 20 different subjects in the Sleep-EDF SC* database.
{See the middle of Figure \ref{FlowChart} for a visualization of each EEG channel by DM.}
In the bottom of Figure \ref{FlowChart}, we plot the embeddings $\{[q_2(i),q_3(i),q_4(i)]\}_{i=1}^J$ and $\{[q_2(i+J),q_3(i+J),q_4(i+J)]\}_{i=1}^J$ for a visualization of {fusing two EEG channels} generated by the multiview DM.
Clearly, we see that Awake, REM, N2 and N3 stages are well clustered in all plots, while N1 is less clustered and tends to mixed up with other stages.
While the sleep dynamics can be easily visualized in Figure \ref{FlowChart}, it is not easy to visualize the temporal dynamics information. For this purpose, we show the final intrinsic features expanded in the time line in Figure \ref{figure4:intrinsic_feature_dynamic}.

Since there are long periods of wakefulness at the start and the end of recordings, when a subject is not sleeping, \cite{DeepSleepNet} only includes 30 minutes of such periods just before and after the sleep periods.
To have a fair comparison, we also follow this truncation rule in this work. In the end, the labeled epochs are imbalanced, with 42.4\% epochs labeled N2 and only 6.6\% epochs labeled N1.

\begin{figure}[!htb]
\subfigure
{\includegraphics[width=.99\textwidth]{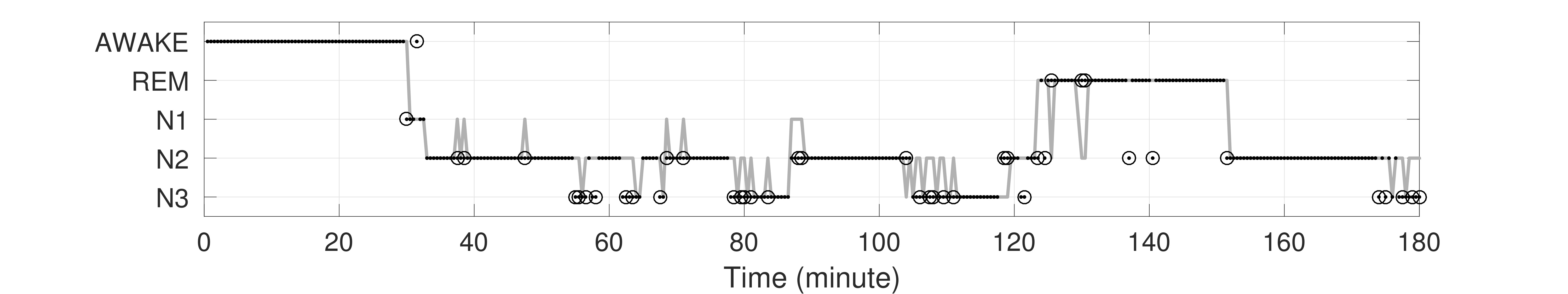}}
\subfigure
{\includegraphics[width=.99\textwidth]{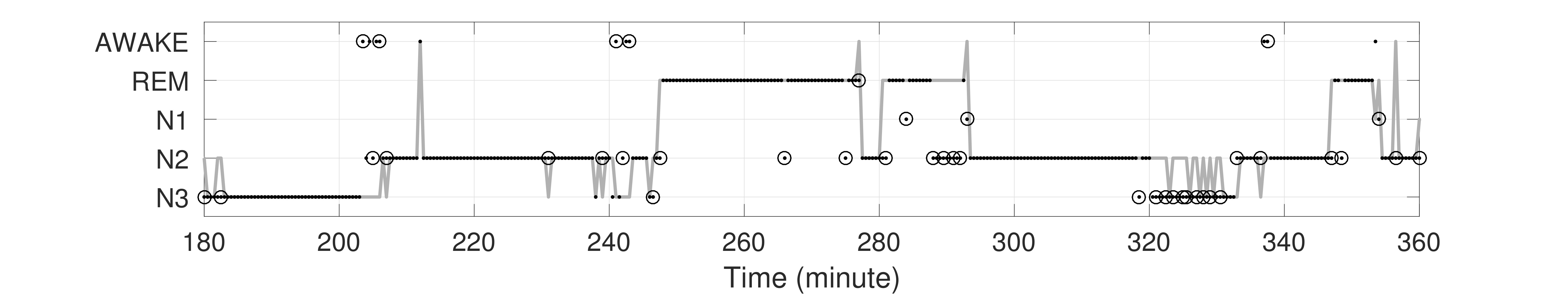}}
\caption{The resulting hypnogram of one subject from SC*. The gray curve is the expert's label, and the black dots are the predicted sleep stages. The discrepancy is emphasized by the black circles.
}
\label{fig:Hypnogram}
\end{figure}

We run the leave-one-subject-out CV.
The averaged confusion matrix of the proposed algorithm over 20 subjects is shown in {Table \ref{table:SC_Fpz-Cz}
for the single-channel case and} Table \ref{table:SC} for the two-channel case. 
{For the the single-channel case (e.g., Fpz-Cz), the overall accuracy is $82.72\%$, the macro F1 is $75.91\%$ and Cohen's kappa equals $76.10\%$.}
For the two-channel case, the overall accuracy is $84.44\%$ and the macro F1 is $78.25\%$, with Cohen's kappa $78.36\%$. Note that the N1 prediction is relatively low compared with other stages. Particularly, the F1 is $46\%$ and most N1 epochs are classified as REM or N2. This misclassification is related to the scattered N1 epochs in Figure \ref{FlowChart} that can be visually observed, and it is the main reason to drag down the overall accuracy and macro F1. We also note that N3 is commonly classified wrongly as N2, Awake is commonly classified wrongly as N1, and REM is commonly classified wrongly as N2.
To further examine the performance, the resulting hypnogram of one subject is shown in  Figure \ref{fig:Hypnogram}. Note that the discrepancy between the experts' annotations and the prediction frequently happens when there is a ``stage transition''. Note that the sleep dynamics transition from one stage to another one often happens in the middle of one epoch. Thus, those epochs with sleep dynamics transition contain information that is not purely for one stage, and hence harder to classify.
In Table \ref{Table:Comparison:EDF-SC}, we compare performance of the proposed algorithm and several reported algorithms validated by the leave-one-subject-out CV scheme.

\begin{table}[h]
\scriptsize
\setlength\extrarowheight{6pt}
\caption{Performance of DM with the scattering transform evaluated by 20-fold leave-one-subject-out cross-validation on the Fpz-Cz channel (top) and Pz-Oz channel (bottom) from the Sleep-EDF SC$^{*}$ database. 
For the Fpz-Cz channel, the overall accuracy equals 82.72\%, the macro F1 score equals 75.91\% and Cohen's kappa equals $76.10\%$. 
The standard deviation of classification accuracy (resp., the macro F1 score and Cohen's kappa) for the 39-night recordings is 8.90\% (resp., 7.71\% and 11.30\%).  
For the Fz-Oz channel, the overall accuracy equals 80.99\%, the macro F1 score equals 72.69\% and
Cohen's kappa equals $73.49\%$. 
The standard deviation of classification accuracy (resp., the macro F1 score and Cohen's kappa) for the 39-night recordings is 5.06\% (resp., 5.99\% and 6.97\%).
The training set for each testing subject consists of the recordings of the remaining 19 subjects during sleep.}
\centering
\begin{tabular}{|c|ccccc|ccc|}
\toprule
\multirow{2}{*}{Fpz-Cz} &\multicolumn{5}{c}{\bf Predicted} & \multicolumn{3}{|c|}{\bf Per-class Metrics}\\
  & \multicolumn{1}{c}{\centering Awake} & \multicolumn{1}{c}{\centering REM} &  \multicolumn{1}{c}{\centering N1} & \multicolumn{1}{c}{\centering N2} & \multicolumn{1}{c}{\centering N3} & \multicolumn{1}{|c}{\centering PR} & \multicolumn{1}{c}{\centering RE} & \multicolumn{1}{c|}{\centering F1} \\
\midrule
Awake (19\%) &  6871 ({\bf 87\%})  &    298   ({\bf 4\%})    &      563 ({\bf 7\%})  &     171  ({\bf2\%})     &      24 ({\bf 0\%}) &  83    & 87  & 85 \\
REM   (18\%) &  295 ({\bf 4\%})    &     6253  ({\bf81\%}) &      378  ({\bf 5\%})  &     790 ({\bf 10\%})    &      1 ({\bf 0\%}) &  79  &  81 &  80 \\
N1    (7\%)  &   549 ({\bf 20\%})   &      616  ({\bf 22\%})  &    1040   ({\bf 37\%})  &      591  ({\bf 21\%})   &     8 ({\bf0\%}) &   46  & 37   &  41 \\
N2    (42\%) &  416 ({\bf 2\%})    &     692  ({\bf 4\%})    &    272    ({\bf 2\%})   &       15821 ({\bf89\%})  &     598 ({\bf3\%}) &   87  &  89  &  88  \\
N3    (14\%) &  101 ({\bf 2\%})    &      13 ({\bf 0\%})      &    0  ({\bf 0\%})       &     872   ({\bf15\%})   &     4717 ({\bf 83\%})  &  88  &  83   &  85  \\
\bottomrule
\end{tabular}

\begin{tabular}{|c|ccccc|ccc|}
\toprule
\multirow{2}{*}{Fz-Oz} &\multicolumn{5}{c}{\bf Predicted} & \multicolumn{3}{|c|}{\bf Per-class Metrics}\\
  & \multicolumn{1}{c}{\centering Awake} & \multicolumn{1}{c}{\centering REM} &  \multicolumn{1}{c}{\centering N1} & \multicolumn{1}{c}{\centering N2} & \multicolumn{1}{c}{\centering N3} & \multicolumn{1}{|c}{\centering PR} & \multicolumn{1}{c}{\centering RE} & \multicolumn{1}{c|}{\centering F1} \\
\midrule
Awake (19\%) &  6803 ({\bf 86\%})  &   420   ({\bf 5\%})    &      466 ({\bf 6\%})  &     220 ({\bf3\%})     &      18 ({\bf 0\%}) &  84    & 86  & 85 \\
REM   (18\%) &  310 ({\bf 4\%})    &     6038   ({\bf78\%}) &      350  ({\bf 5\%})  &     1019  ({\bf 13\%})    &      0 ({\bf 0\%}) &  77  &  78 &  78 \\
N1    (7\%)  &   794 ({\bf 28\%})   &      647  ({\bf 23\%})  &    755   ({\bf 27\%})  &      602  ({\bf 22\%})   &     6 ({\bf0\%}) &   42  & 27   &  33 \\
N2    (42\%) &  211 ({\bf 1\%})    &    703  ({\bf 4\%})    &    226    ({\bf 1\%})   &       15943 ({\bf90\%})  &     716 ({\bf4\%}) &   84  &  90  &  87  \\
N3    (14\%) &  4 ({\bf 0\%})    &      9 ({\bf 0\%})      &    2  ({\bf 0\%})       &     1250   ({\bf22\%})   &     4438 ({\bf 78\%})  &  86 &  78   &  82  \\
\bottomrule
\end{tabular}

\label{table:SC_Fpz-Cz}
\end{table}

\begin{table}
\scriptsize
\setlength\extrarowheight{3pt}
\caption{Performance of the multiview DM with the scattering transform evaluated by 20-fold  leave-one-subject-out  cross-validation on the
Fpz-Cz and Pz-Oz  channels  from  the Sleep-EDF SC$^{*}$
database. The overall accuracy equals 84.44\%, the macro F1 score equals 78.25\% and
Cohen's kappa equals $78.36\%$.  If the classification accuracy, macro F1 score, and Cohen's kappa are computed for each night recording, the standard deviation of classification accuracy (resp., the macro F1 score and Cohen's kappa) for the 39-night recordings is 5.36\% (resp., 6.40\% and 8.01\%).  The training set for each testing subject consists of the recordings of the remaining 19 subjects during sleep.}
\centering
\begin{tabular}{|c|ccccc|ccc|}
\toprule
\multirow{2}{*}{Fpz-Cz+Pz-Oz} &\multicolumn{5}{c}{\bf Predicted} & \multicolumn{3}{|c|}{\bf Per-class Metrics}\\
  & \multicolumn{1}{c}{\centering Awake} & \multicolumn{1}{c}{\centering REM} &  \multicolumn{1}{c}{\centering N1} & \multicolumn{1}{c}{\centering N2} & \multicolumn{1}{c}{\centering N3} & \multicolumn{1}{|c}{\centering PR} & \multicolumn{1}{c}{\centering RE} & \multicolumn{1}{c|}{\centering F1} \\
\midrule
Awake (19\%) &  7034 ({\bf 89\%})  &    148   ({\bf 2\%})    &      525 ({\bf 7\%})  &     197  ({\bf2\%})     &      23 ({\bf 0\%}) &  90    & 89  & 90 \\
REM   (18\%) &  125 ({\bf 2\%})    &     6070  ({\bf79\%}) &      528  ({\bf 7\%})  &     991 ({\bf 13\%})    &      3 ({\bf 0\%}) &  85  &  79 &  82 \\
N1    (7\%)  &   498 ({\bf 18\%})   &      436  ({\bf 16\%})  &    1218   ({\bf 43\%})  &      643  ({\bf 23\%})   &     9 ({\bf0\%}) &   47  & 43   &  46 \\
N2    (42\%) &  115 ({\bf 1\%})    &     492 ({\bf 3\%})    &    313    ({\bf 2\%})   &       16337 ({\bf92\%})  &     542 ({\bf3\%}) &   86  &  92  &  89  \\
N3    (14\%) &  17 ({\bf 0\%})    &      0 ({\bf 0\%})      &    1  ({\bf 0\%})       &     921   ({\bf16\%})   &     4764 ({\bf 84\%})  &  89  &  86   &  84  \\
\bottomrule
\end{tabular}
\label{table:SC}
\end{table}

\begin{table}
\setlength\extrarowheight{8pt}
\caption{Comparison between DM-HMM and other sleep stage scoring methods in terms of the overall accuracy (ACC), macro-F1 score (MF1) and Cohen's kappa. 
{Note that the total number of epochs in the ST$^*$ dataset is $21076$ but the performance of our algorithm is evaluated on $20988$ epochs. In fact, we do not predict the sleep stages of the first $3$ epochs and the last epoch of each recording in the ST$^*$ dataset because the features for each epoch is extracted from a $90$s signal. 
As a result, $21076-88=20988$ epochs in the ST$^*$ dataset are evaluated.}
\label{Table:Comparison:EDF-SC}}
\rotatebox{90}{
\begin{tabular}{ccccc}
\toprule
Author & EEG Channel & Extracted features & Model$\setminus$Algorithm & ACC$\setminus$MF1\\
\midrule
\cite{DeepSleepNet}  & \multicolumn{1}{V{8em}}{\centering{\bf SC$^{*}$ Fpz-Cz}\\ (41950 epochs)
\\ - - - - - - - - - -  \\ {\bf SC$^{*}$ Pz-Oz}\\ (41950 epochs)}
& \multicolumn{1}{V{15em}}{Raw data without any prior processing}  & \multicolumn{1}{V{9em}}{Recurrent Neural Networks with class-balanced random sampling and stochastic gradient {descent}}     & \multicolumn{1}{V{6em}}{\centering82.0\%$\setminus$76.9\%\\ {($\kappa$=76\%)}
\\ - - - - - - - - - -  \\79.8\%$\setminus$73.1\%\\ ($\kappa$=72\%)}
\\
\hline
\cite{2016autoencoder} & \multicolumn{1}{V{8em}}{\centering {\bf SC$^{*}$ Fpz-Cz}\\ (37022 epochs)} & \multicolumn{1}{V{17em}}{(a) Complex Morlet wavelets  \\(b) Time-domain amplitude characteristics \\(c) Pearson correlation coefficient between each pair of frequency-band signals \\(d) Autocorrelation in the time domain signal for 0.5s lags} &
\multicolumn{1}{V{9em}}{Stacked sparse autoencoders induced by
the class-balanced random sampling}  &  \multicolumn{1}{V{6em}}{\centering 78.9\%$\setminus$73.7\%}\\
\hline
\cite{tsinalis2016automatic} & \multicolumn{1}{V{8em}}{\centering {\bf SC$^{*}$ Fpz-Cz}\\ (37022 epochs)} & \multicolumn{1}{V{15em}}{Raw data without any prior processing}  &
 \multicolumn{1}{V{9em}}{Convolutional Neural Networks with class-balanced random sampling and stochastic gradient descend}  & \multicolumn{1}{V{6em}}{\centering 74.8\%$\setminus$69.8\%}\\
\hline
\multirow{9}{*}{{\bf Ours} }              & \multicolumn{1}{V{9em}}{
\centering {\bf SC$^{*}$ Fpz-Cz}\\ (41950 epochs)\\
- - - - - - - - - - - - - - - - - \\
\centering {\bf SC$^{*}$ Pz-Oz}\\ (41950 epochs)\\
- - - - - - - - - - - - - - - - - \\
\centering {\bf SC$^{*}$ Fpz-Cz + Pz-Oz}\\ (41950 epochs)\\
- - - - - - - - - - - - - - - - - \\
\centering {\bf ST$^{*}$ Fpz-Cz}\\ (20988 epochs)\\
- - - - - - - - - - - - - - - - -\\
\centering {\bf ST$^{*}$ Pz-Oz}\\ (20988 epochs)\\
- - - - - - - - - - - - - - - - - \\
\centering {\bf ST$^{*}$ Fpz-Cz + Pz-Oz}\\ (20988 epochs)}&
\multicolumn{1}{V{15em}}{\centering Scattering spectrum \\ (multiview) diffusion map} & \multicolumn{1}{V{9em}}{Support Vector Machine}    &\multicolumn{1}{V{6em}}{
82.72\%$\setminus$75.91\%\\ \centering($\kappa$=76.10\%)
\\- - - - - - - - - -  \\
80.99\%$\setminus$72.69\%\\ \centering($\kappa$=73.49\%)
\\- - - - - - - - - -  \\
84.44\%$\setminus$78.25\%\\ \centering($\kappa$=78.36\%)
\\- - - - - - - - - -  \\
78.63\%$\setminus$73.58\%\\ \centering($\kappa$=69.48\%)
\\- - - - - - - - - -  \\
75.74\%$\setminus$69.97\%\\ \centering($\kappa$=65.39\%)
\\- - - - - - - - - -  \\
79.05\%$\setminus$74.73\%\\ \centering($\kappa$=70.31\%)}\\
\bottomrule
\end{tabular}
}
\end{table}

\subsection{Sleep-EDF database (ST*)}

\begin{figure}[!htb]
\centering
\includegraphics[width=.49\textwidth]{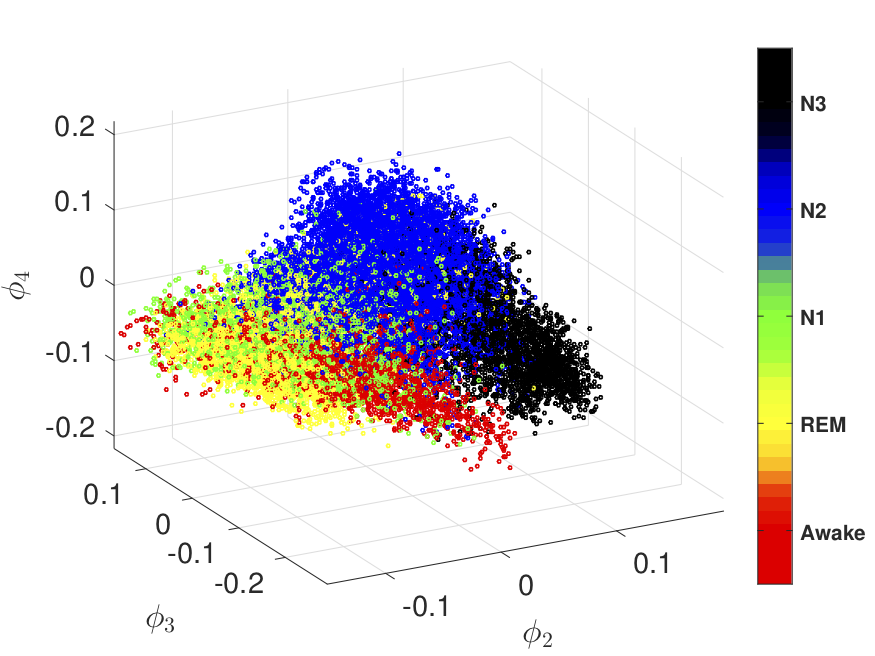}\label{Figure:2:DM1_ST}
\includegraphics[width=.49\textwidth]{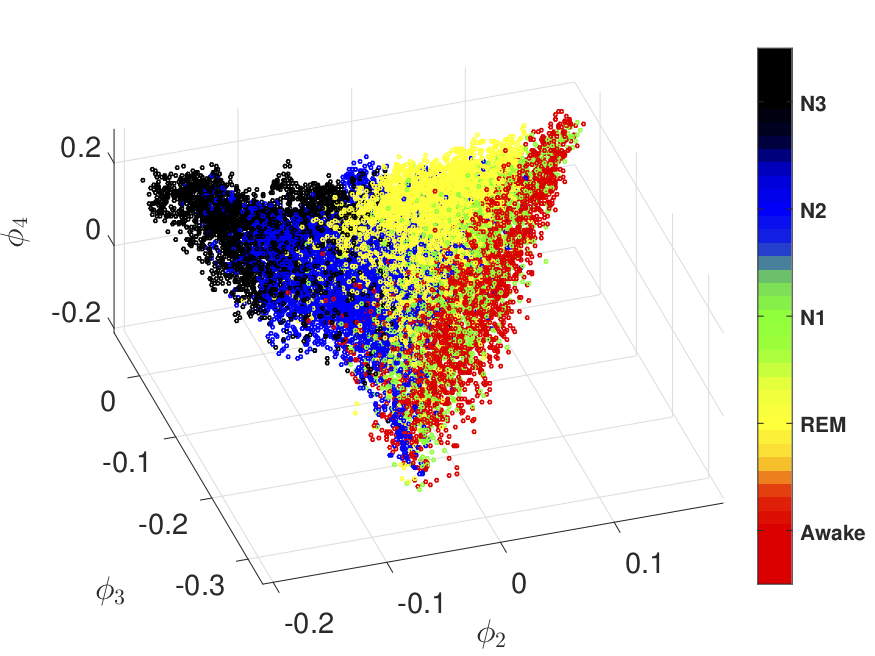}\label{Figure:2:DM2_ST}
\caption{A visualization of the intrinsic sleep features (from single channel)  extracted from 22 different subjects from the Sleep-EDF database (ST*).
In subplot \ref{Figure:2:mvDM1_ST}, we plot $\{[q_2(i),q_3(i),q_4(i)]\}_{i=1}^J$, and in subplot \ref{Figure:2:mvDM2_ST}, we show $\{[q_2(i+J),q_3(i+J),q_4(i+J)]\}_{i=1}^J$.
The ratios of the stages Awake, REM, N1, N2, and N3 are
10\%, 20\%, 10\%, 45\%, and 15\% respectively. Each point corresponds to a 30-second epoch.
}
\label{fig:STvisulizationDM}
\end{figure}

\begin{figure}[!htb]
\centering
\includegraphics[width=.49\textwidth]{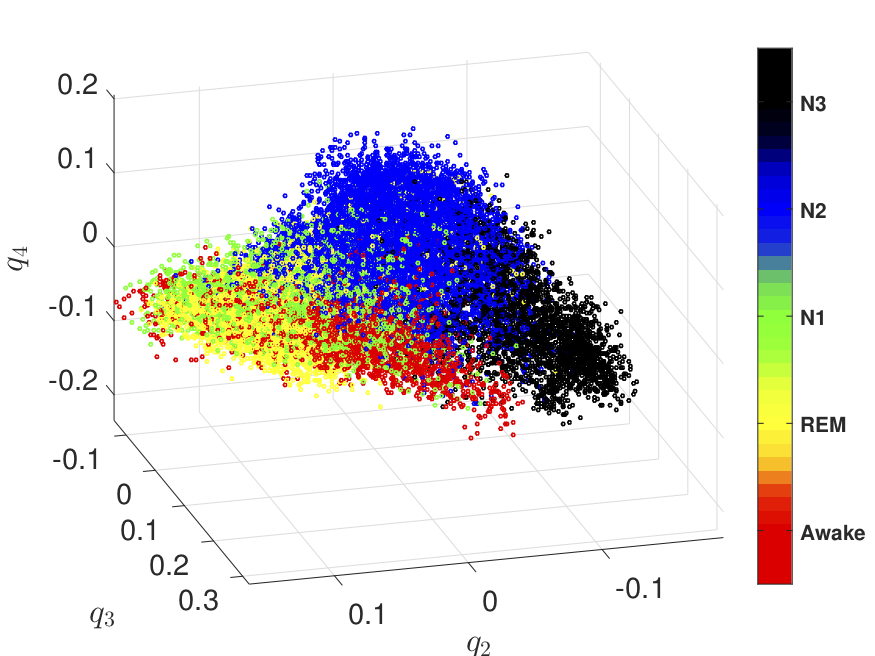}\label{Figure:2:mvDM1_ST}
\includegraphics[width=.49\textwidth]{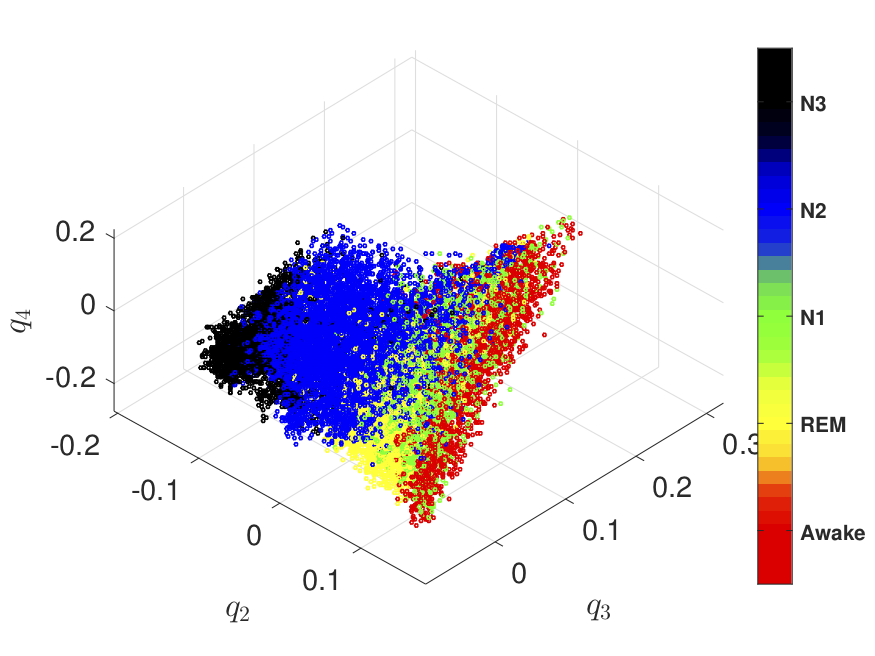}\label{Figure:2:mvDM2_ST}
\caption{A visualization of the common intrinsic sleep features (from two channels) extracted from 22 different subjects from the Sleep-EDF database (ST*).
In subplot \ref{Figure:2:mvDM1_ST}, we plot $\{[q_2(i),q_3(i),q_4(i)]\}_{i=1}^J$, and in subplot \ref{Figure:2:mvDM2_ST}, we show $\{[q_2(i+J),q_3(i+J),q_4(i+J)]\}_{i=1}^J$.
The ratios of the stages Awake, REM, N1, N2, and N3 are
10\%, 20\%, 10\%, 45\%, and 15\% respectively. Each point corresponds to a 30-second epoch.
}
\label{fig:STvisulization}
\end{figure}

See Figures \ref{fig:STvisulizationDM} and \ref{fig:STvisulization} for a visualization of DM and multiview DM of the ST* database. While Awake, REM, N2 and N3 stages are still well clustered in all plots, compared with the normal subjects in the SC* database, the separation is less clear.

\begin{table}[h]
\scriptsize
\setlength\extrarowheight{6pt}
\caption{Performance of DM with the scattering transform evaluated by 22-fold leave-one-subject-out cross-validation on the Fpz-Cz or Pz-Oz channel from the Sleep-EDF ST$^{*}$ database. 
For Fpz-Cz, the overall accuracy equals 78.63\%, the macro F1 score equals 73.58\% and
Cohen's kappa equals $69.48\%$. The standard deviation of classification accuracy (resp., the macro F1 score and Cohen's kappa) for the 22-night recordings is 7.97\%
(resp., 8.11\% and 10.25\%).
For Pz-Oz, the overall accuracy equals 75.74\%, the macro F1 score equals 69.97\% and
Cohen's kappa equals $65.39\%$. The standard deviation of classification accuracy (resp., the macro F1 score and Cohen's kappa) for the 22-night recordings is 7.24\%
(resp., 8.30\% and 8.92\%).
}
\centering
\begin{tabular}{|c|ccccc|ccc|}
\toprule
 \multirow{2}{*}{Fpz-Cz} &\multicolumn{5}{c}{\bf Predicted} & \multicolumn{3}{|c|}{\bf Per-class Metrics}\\
  & \multicolumn{1}{c}{\centering Awake} & \multicolumn{1}{c}{\centering REM} &  \multicolumn{1}{c}{\centering N1} & \multicolumn{1}{c}{\centering N2} & \multicolumn{1}{c}{\centering N3} & \multicolumn{1}{|c}{\centering PR} & \multicolumn{1}{c}{\centering RE} & \multicolumn{1}{c|}{\centering F1} \\
\midrule
Awake (10\%) &  1726 ({\bf 78\%})  &    63   ({\bf 3\%})    &      364 ({\bf 16\%})  &     55  ({\bf3\%})     &      5 ({\bf 0\%}) &  79   & 78  & 78 \\
REM   (20\%) &  84 ({\bf 2\%})    &     3196  ({\bf77\%}) &      326  ({\bf 8\%})  &     518 ({\bf 13\%})    &      6 ({\bf 0\%}) &  86  &  77 &  82 \\
N1    (10\%)  &   327 ({\bf 16\%})   &      256  ({\bf 13\%})  &    1003   ({\bf 49\%})  &      444  ({\bf 22\%})   &     2 ({\bf0\%}) &   50  & 49   &  50 \\
N2    (45\%) &  51 ({\bf 0\%})    &     190 ({\bf 2\%})    &    307    ({\bf 3\%})   &       8502 ({\bf90\%})  &     439 ({\bf5\%}) &   81  &  90  &  85 \\
N3    (15\%) &  10 ({\bf 0\%})    &      1 ({\bf 0\%})      &    7  ({\bf 0\%})       &     1034   ({\bf33\%})   &     2093 ({\bf 67\%})  &  82  &  67   &  74  \\
\bottomrule
\end{tabular}
\label{table:ST_Fpz-Cz}

\begin{tabular}{|c|ccccc|ccc|}
\toprule
 \multirow{2}{*}{Pz-Oz} &\multicolumn{5}{c}{\bf Predicted} & \multicolumn{3}{|c|}{\bf Per-class Metrics}\\
  & \multicolumn{1}{c}{\centering Awake} & \multicolumn{1}{c}{\centering REM} &  \multicolumn{1}{c}{\centering N1} & \multicolumn{1}{c}{\centering N2} & \multicolumn{1}{c}{\centering N3} & \multicolumn{1}{|c}{\centering PR} & \multicolumn{1}{c}{\centering RE} & \multicolumn{1}{c|}{\centering F1} \\
\midrule
Awake (10\%) &  1803 ({\bf 81\%})  &    38   ({\bf 2\%})    &      334 ({\bf 15\%})  &     37  ({\bf2\%})     &      1 ({\bf 0\%}) &  74   & 81  & 78 \\
REM   (20\%) &  59 ({\bf 2\%})    &     3272  ({\bf79\%}) &      204  ({\bf 5\%})  &     592 ({\bf 14\%})    &      3 ({\bf 0\%}) &  78  & 79 &  79 \\
N1    (10\%)  &   449 ({\bf 22\%})   &      383  ({\bf 19\%})  &    801   ({\bf 39\%})  &      395  ({\bf 20\%})   &     4 ({\bf0\%}) &   49  & 39   &  44 \\
N2    (45\%) &  110 ({\bf 1\%})    &     401 ({\bf 4\%})    &    282    ({\bf 3\%})   &       8176 ({\bf86\%})  &     520 ({\bf6\%}) &   79  &  86  &  82 \\
N3    (15\%) &  5 ({\bf 0\%})    &      75 ({\bf 3\%})      &    9  ({\bf 0\%})       &     1195   ({\bf38\%})   &     1861 ({\bf 59\%})  &  78  &  59   &  67  \\
\bottomrule
\end{tabular}

\end{table}

\begin{table}
\scriptsize
\setlength\extrarowheight{6pt}
\caption{Performance of the multiview DM with the scattering transform evaluated by 22-fold leave-one-subject-out cross-validation on the Fpz-Cz and Pz-Oz channels from the Sleep-EDF ST$^{*}$ database. The overall accuracy equals 79.05\%, the macro F1 score equals 74.73\% and
Cohen's kappa equals $70.31\%$. The standard deviation of classification accuracy (resp., the macro F1 score and Cohen's kappa) for the 22-night recordings is 7.21\%
(resp., 7.58\% and 9.38\%).}
\centering
\begin{tabular}{|c|ccccc|ccc|}
\toprule
\multirow{2}{*}{Fpz-Cz+Pz-Oz} &\multicolumn{5}{c}{\bf Predicted} & \multicolumn{3}{|c|}{\bf Per-class Metrics}\\
  & \multicolumn{1}{c}{\centering Awake} & \multicolumn{1}{c}{\centering REM} &  \multicolumn{1}{c}{\centering N1} & \multicolumn{1}{c}{\centering N2} & \multicolumn{1}{c}{\centering N3} & \multicolumn{1}{|c}{\centering PR} & \multicolumn{1}{c}{\centering RE} & \multicolumn{1}{c|}{\centering F1} \\
\midrule
Awake (10\%) &  1838 ({\bf 83\%})  &    19   ({\bf 1\%})    &      323 ({\bf 15\%})  &     29  ({\bf1\%})     &      4 ({\bf 0\%}) &  79   & 83  & 81 \\
REM   (20\%) &  26 ({\bf 1\%})    &     3344  ({\bf81\%}) &      288  ({\bf 7\%})  &     465 ({\bf 11\%})    &      7 ({\bf 0\%}) &  86  &  81 &  84 \\
N1    (10\%)  &   344 ({\bf 17\%})   &      244  ({\bf 12\%})  &    1059   ({\bf 52\%})  &      380  ({\bf 19\%})   &     5 ({\bf0\%}) &   52  & 52   &  52 \\
N2    (45\%) &  109 ({\bf 1\%})    &     253 ({\bf 3\%})    &    352    ({\bf 4\%})   &       8265 ({\bf87\%})  &     510 ({\bf5\%}) &   81  &  87  &  84 \\
N3    (15\%) &  8 ({\bf 0\%})    &      10 ({\bf 0\%})      &    10  ({\bf 0\%})       &     1016   ({\bf32\%})   &     2101 ({\bf 67\%})  &  80  &  67   &  73  \\
\bottomrule
\end{tabular}
\label{table:ST_twochannel}
\end{table}
We run the leave-one-subject-out CV.
The averaged confusion matrix of the proposed algorithm over 22 subjects is shown {in Table \ref{table:ST_Fpz-Cz}
for the single-channel case} and Table \ref{table:ST_twochannel} for the two-channel case.
{For the single channel case (e.g., Fpz-Cz channel), the overall accuracy is $78.63\%$, the macro F1 is $73.58\%$, and Cohen's kappa is $69.48\%$}.
For the two-channel case, the overall accuracy is $79.05\%$ and the macro F1 is $74.73\%$, with Cohen's kappa $70.31\%$.
In this database, there are $10\%$ epochs labeled as N1, which is slightly higher than that of the SC$^{*}$ database. While the prediction performance of N1 is slightly higher ($50\%$ when we use the Fpz-Cz channel and $52\%$ when we use two channels), it is still relatively low. Also, note that the prediction performance of N3 is lower, and a significant portion of N3 is mis-classified as N2.

\subsection{More comparisons}

\subsubsection{Handling imbalanced data}
An ideal approach to handle the imbalanced data is collecting more data {for the small-sized group} to enhance the prediction accuracy. In the SC${}^*$ dataset,
there were long periods of awake epochs before the start and after the end of sleep that we can use. To further evaluate the algorithm, we consider longer periods of wakefulness just before and after the sleep periods.
Apart from the {2 recordings (sc4092e0 and sc4192e0)}, we included 90 minutes of awake periods before and after the sleep periods.
For the {sc4092e0 and sc4192e0} recordings, we only included 60 minutes of awake periods just before and after the sleep periods due to the appearance of artifacts (labeled as MOVEMENT and UNKNOWN), which were in the start or the end of each recording. With more awake epochs, the corresponding comparison matrix is shown in
Table \ref{table:SC_moreAwake}. 
{ In comparison with the result subject to the 30-minute truncation rule, the accuracy and F1 of Awake increase from 88\% and 90\% to 91\% and 94\% respectively, and the accuracy and F1 of N1 change from 43\% and 46\% to 63\% and 43\% respectively. Note that the F1 of N1 decreases since the precision decreases due to the increased number of awake epochs and increased number of epochs erroneously classified as N1. 

Another approach is handling the imbalanced data according to the class-balanced random sampling scheme proposed in \cite{2016autoencoder} under the same 30-minute truncation rule. 
Specifically, for each recording, we randomly sample $K$ epochs for per-stage, where $K$ is the number of epochs of the least represented stage (N1). 
We apply this class-balanced random sampling scheme before using the SVM to train the common intrinsic sleep features. See Table \ref{table:SC_moreAwake} for details. Compared with Table \ref{table:SC}, the accuracy of N1 can be improved from 43\% to 76\% and the overall accuracy decreases to 81.12\%.
}

\begin{table}[!htb]
\scriptsize
\setlength\extrarowheight{3pt}
\caption{Performance of the multiview DM with the scattering transform evaluated by 20-fold leave-one-subject-out cross-validation on Fpz-Cz and Pz-Oz channels from the Sleep-EDF SC* database with two different approaches to handle the imbalanced groups. (1) take a longer awake periods; (2) apply the class-balanced random sampling scheme proposed in \cite{2016autoencoder}. 
For the approach (1), the overall accuracy equals 83.69\%, the macro F1 score equals 77.11\% and Cohen's kappa equal 78.14\%. If the classification accuracy, macro F1 score, and Cohen's kappa are computed for each night recording, the standard deviation of classification accuracy (resp. the macro F1 score and Cohen's kappa) for the 39-night recordings is 7.23\% (resp. 5.47\% and 8.77\%).
For the approach (2), the overall accuracy equals 81.12\%, the macro F1 score equals 77.33\% and
Cohen's kappa equals $74.82\%$. If the classification accuracy, macro F1 score, and Cohen's kappa are computed for each night recording, the standard deviation of classification accuracy (resp., the macro F1 score and Cohen's kappa) for the 39-night recordings is 6.29\% (resp., 7.30\% and 8.77\%).  
}
\centering
\begin{tabular}{|c|ccccc|ccc|}
\toprule
\multicolumn{9}{|c|}{\bf Longer awake periods} \\
\hline
\multirow{2}{*}{Fpz-Cz+Pz-Oz} &\multicolumn{5}{c}{\bf Predicted} & \multicolumn{3}{|c|}{\bf Per-class Metrics}\\
  & \multicolumn{1}{c}{\centering Awake} & \multicolumn{1}{c}{\centering REM} &  \multicolumn{1}{c}{\centering N1} & \multicolumn{1}{c}{\centering N2} & \multicolumn{1}{c}{\centering N3} & \multicolumn{1}{|c}{\centering PR} & \multicolumn{1}{c}{\centering RE} & \multicolumn{1}{c|}{\centering F1} \\
\midrule
Awake ({34}\%) &  15478 ({\bf 91\%})  &    221   ({\bf 1\%})    &    1245 ({\bf 7\%})  &    76  ({\bf1\%})     &    44 ({\bf 0\%}) &  97    & 91  & 94 \\
REM   ({15}\%) &  114 ({\bf 2\%})    &     5954  ({\bf77\%}) &      1071  ({\bf 14\%})  &    570  ({\bf 7\%})    &    8 ({\bf 0\%}) &  80  &  77 &  79 \\
N1    ({5}\%)  &   167 ({\bf 6\%})   &      499  ({\bf 18\%})  &    1754   ({\bf 63\%})  &      351  ({\bf 12\%})   &   33 ({\bf1\%}) &   33  & 63   &  43 \\
N2    ({35}\%) &  130 ({\bf 1\%})    &     740  ({\bf 4\%})    &    1196    ({\bf 6\%})   &       14337 ({\bf81\%})  &   1396 ({\bf8\%}) &   91 &  81  &  85  \\
N3    ({11}\%) &  18 ({\bf 0\%})    &      3 ({\bf 0\%})      &   44  ({\bf 1\%})       &     407   ({\bf7\%})   &   5231 ({\bf 92\%})  &  78 &  92   &  84  \\
\bottomrule
\end{tabular}
\begin{tabular}{|c|ccccc|ccc|}
\toprule
\multicolumn{9}{|c|}{\bf Class-balanced random sampling scheme \cite{2016autoencoder}} \\
\hline
\multirow{2}{*}{Fpz-Cz+Pz-Oz}  &\multicolumn{5}{c}{\bf Predicted} & \multicolumn{3}{|c|}{\bf Per-class Metrics}\\
  & \multicolumn{1}{c}{\centering Awake} & \multicolumn{1}{c}{\centering REM} &  \multicolumn{1}{c}{\centering N1} & \multicolumn{1}{c}{\centering N2} & \multicolumn{1}{c}{\centering N3} & \multicolumn{1}{|c}{\centering PR} & \multicolumn{1}{c}{\centering RE} & \multicolumn{1}{c|}{\centering F1} \\
\midrule
Awake (19\%) &  6444 ({\bf 81\%})  &    103   ({\bf 1\%})    &    1313 ({\bf 17\%})  &    31  ({\bf0\%})     &      36 ({\bf 1\%}) &  96    & 81 & 88 \\
REM   (18\%) &  35 ({\bf 1\%})    &     5794  ({\bf75\%}) &     1322  ({\bf 17\%})  &     563 ({\bf 7\%})    &      3 ({\bf 0\%}) &  87  &  75 &  81 \\
N1    (7\%)  &   186 ({\bf 7\%})   &      238  ({\bf 8\%})  &    2123  ({\bf 76\%})  &     242  ({\bf 9\%})   &     15 ({\bf0\%}) &   33  & 76   &  46 \\
N2    (42\%) &  58 ({\bf 0\%})    &     529 ({\bf 3\%})    &    1595   ({\bf 9\%})   &       14517 ({\bf82\%})  &     1100 ({\bf6\%}) &  92  &  82  &  86  \\
N3    (14\%) &  15 ({\bf 0\%})    &      1 ({\bf 0\%})      &    65 ({\bf 1\%})       &     469  ({\bf8\%})   &     5153 ({\bf 91\%})  &  82  &  90   &  86  \\
\bottomrule
\end{tabular}
\label{table:SC_moreAwake}
\end{table}

\subsubsection{With and without the sensor fusion}

To appreciate the significance of the diffusion geometry based sensor fusion framework, we report the results without the critical setup in the proposed algorithm -- the sensor fusion. 
We consider the case if we simply concatenate intrinsic sleep features of two channels, instead of taking the common intrinsic sleep features; that is, we concatenate $\Phi^x_t(\mathbf{u}^{(j)})$ and $\Phi^y_t(\mathbf{u}^{(j)})$ in \eqref{DM} directly to replace \eqref{Definition:common intrinsic sleep feature} when we train the {kernel SVM} model.
We also compare the single EEG channel result with the fusion result; that is, we train the {kernel SVM} on the intrinsic sleep features extracted from the Fpz-Cz or Pz-Oz channel.

\begin{figure}[!htb]
\centering
\subfigure
{\includegraphics[width=.9\textwidth]{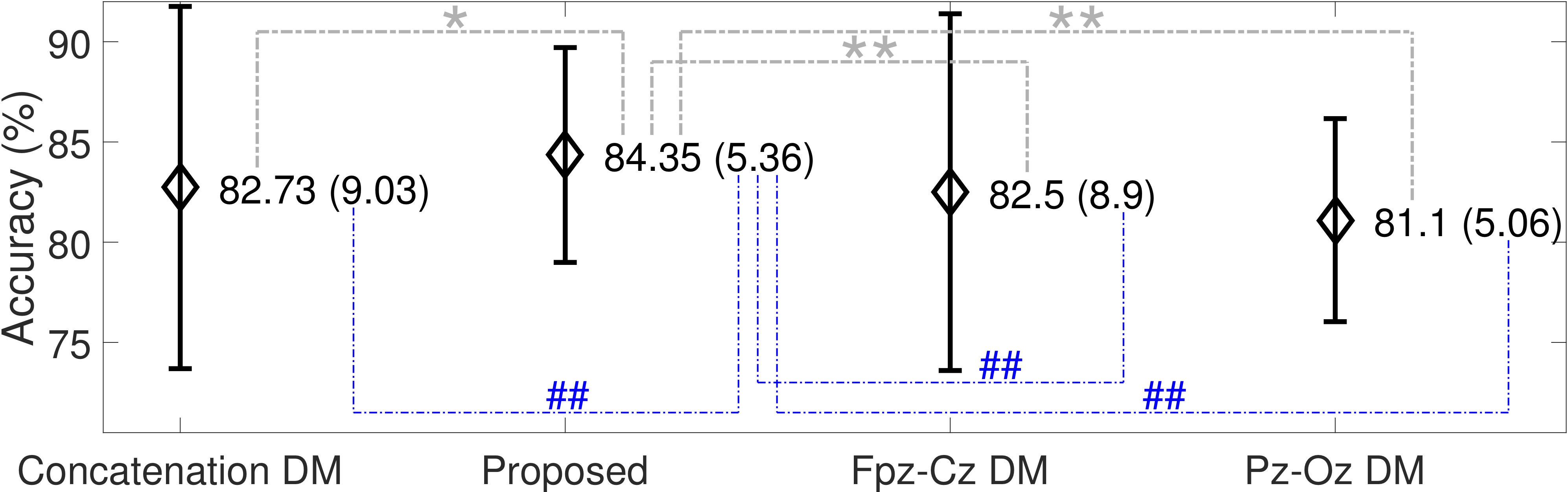}}
\subfigure
{\includegraphics[width=.9\textwidth]{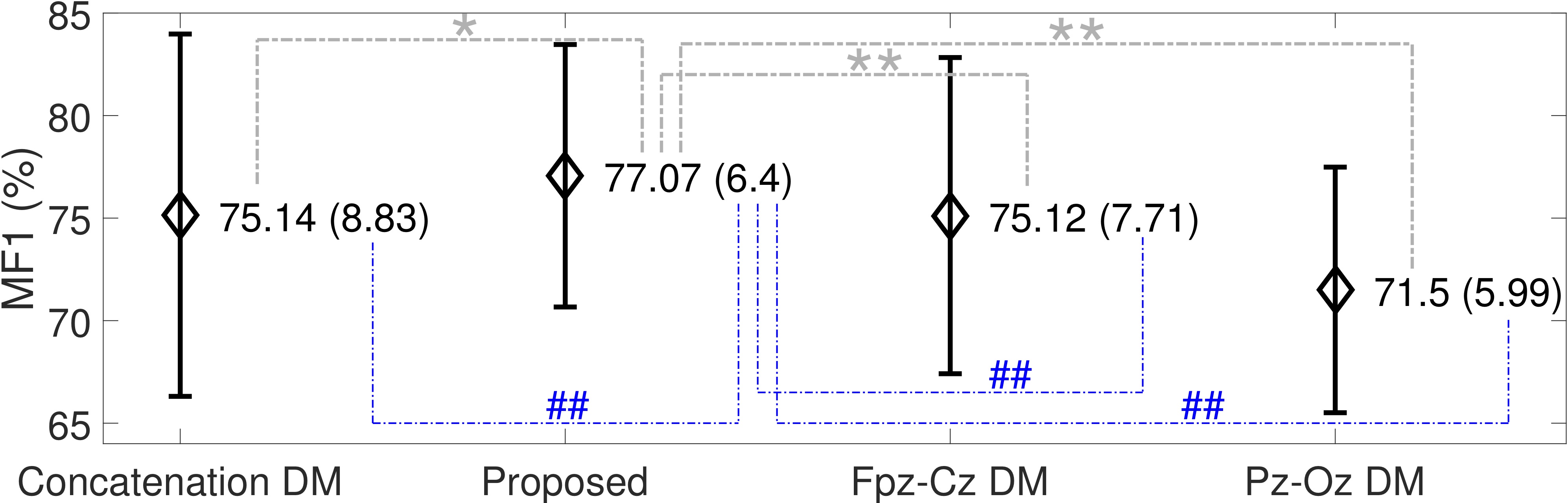}}
\subfigure
{\includegraphics[width=.9\textwidth]{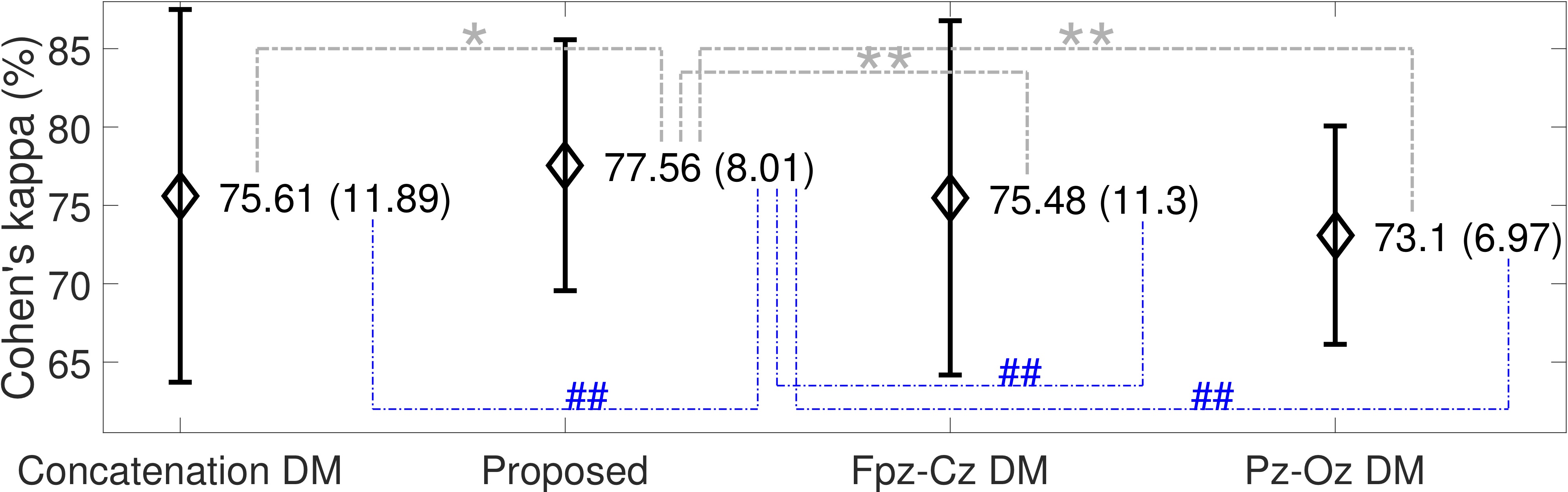}}
\caption{Comparison between different information fusion methods in terms of the accuracy
(ACC), macro F1-score (MF1) and Cohen's Kappa for the SC* database. 
To evaluate if the
mean is improved, the one-tail Wilcoxon signed-rank test is applied under the null hypothesis
that the difference between the pairs follows a symmetric distribution around zero. 
To evaluate
if the variance is smaller, we apply the one-tail F test under the null hypothesis that there is
no difference between the variances. 
$\star$ (respectively $\star\star$) means statistical significance without
(respectively with) the Bonferroni connection when the mean is compared; 
\# (respectively
\#\#) means statistical significance without (respectively with) the Bonferroni connection when the variance is compared.
}
\label{fig:effect_AD}
\end{figure}

Different comparisons of the SC${}^*$ database are shown in Figure \ref{fig:effect_AD}. 
It is clear that the averaged ACC, MF1 and Cohen's kappa are consistently downgraded in these three cases.
In Figure \ref{fig:effect_AD}, we see that compared with single channel or sensor fusion with a direct concatenation, the proposed sensor fusion of two channels consistently improve the result with statistical significance, for both mean and variance of ACC, MF1 and Cohen's kappa.  
This fact reflects the essential property of the diffusion-based algorithms. Via the diffusion-based sensor fusion, the intrinsic sleep features are ``stabilized'', and hence a smaller variance. 

\section{Discussion and Conclusion}

The diffusion geometry based sensor fusion framework is proposed to capture the geometric structure of the sleep dynamics. We take the {scattering transform} of EEG signals as the initial feature and test the framework on the publicly available benchmark database. 
With the learning algorithm {SVM}, we obtain an accurate prediction model, and the result is compatible with several state-of-the-art algorithms based on neural network (NN). In addition, the chosen tools all have solid theoretical backup. All these summarize the usefulness of the diffusion geometry framework for the sleep dynamics study. We mention that the proposed framework is flexible to study other physiological dynamics but not only for studying the sleep dynamics. For example, its variation has been applied to study f-wave from subjects with atrial fibrillation \cite{Malik_Reed_Wang_Wu:2017}, intra-cranial EEG signal \cite{Alagapan_Shin_Frohlich_Wu:2018}, fetal electrocardiogram (ECG) extraction from single lead trans-abdominal maternal ECG signal \cite{Li_Frasch_Wu:2017}, etc.

\subsection{Dealing with misclassification of stage 1 sleep}
Although our overall prediction accuracy is compatible with the state-of-the-art prediction algorithm in the Sleep-EDF SC* database, like \cite{DeepSleepNet}, we see that the prediction accuracy of N1 is low by our algorithm {(when there is only a single channel, Fpz-Cz, and with the same number of epochs, the F1 of N1 prediction is 41\% by our method and 46.6\% in \cite[Table III]{DeepSleepNet}, and the overall accuracy is 82.72\% by our method and 82\% in \cite[Table III]{DeepSleepNet})}. This prediction rate of N1 is also low in the Sleep-EDF ST* database. This low prediction rate partially comes from the relatively small size of N1 epochs, and partially comes from available channels {that we discuss below.} 

Based on the AASM criteria \cite{Iber2007,berry2012aasm}, to distinguish N1 and REM, we need electrooculogram and electromyogram {signals}, which are not available in the dataset. The EEG backgrounds of N1 and N2 are the same, and experts distinguish N1 and N2 by the K-complex or spindle, as well as the {\em 3-minute rule}.
{While we use 90 seconds signal to establish intrinsic sleep features,} the 3-minute rule is not considered in the algorithm. 
In the proposed algorithm, the temporal information among epochs is not fully utilized when we design the intrinsic sleep feature.
How to incorporate the temporal information into the diffusion geometry framework will be explored in the future. 
Furthermore, there are other information in addition to the spectral information discussed in this paper. We do not extensively explore all possible information, but focus on the diffusion geometry and sensor fusion framework. For example, while the vertex sharp is a common ``landmark'' indicating transition from N1 to N2, we do not take it into account. 
Another interesting reasoning that it is possible to improve the N1 accuracy is based on the deep neural network (DNN) result \cite{DeepSleepNet}. This suggests that by taking experts' labels into account, some distinguishable EEG structure of N1 that is not sensible {the scattering transform} can be efficiently extracted by the DNN framework proposed in \cite{DeepSleepNet}. In conclusion, since the proposed features depend solely on the {scattering transform}, we may need features of different categories to capture this unseen N1 structure.

Note that beside N1, the prediction performance of N3 is also lower in the Sleep-EDF ST* database, where the subjects take temazepam before data recording. It has been well known that in general benzodiazepine hypnotics \cite{Borbely1985} reduces the low frequency activity and enhances spindle frequency. Since our features are mainly based on the spectral information, a N3 epoch might look more like N2 epochs, and hence the confusion and the lower performance. This outcome emphasizes the importance of the drug history when {designing} the algorithm.

\subsection{Comparison with DNN}
Compared with the DNN approach \cite{DeepSleepNet}, which is supervised in nature, our {feature extraction step} is unsupervised in nature. Recall that the main difference between the supervised learning and unsupervised learning is that the label information is taken into account in the supervised learning approach. The success of DNN in many fields is well known \cite{LeCun2015}, and it is not surprised that it has a great potential to help medical data analysis. 

While DNN is in general an useful tool for the engineering purpose, it is often criticized of working as a black box. For medical problems and datasets, when interpretation is needed, a mathematically solid and interpretable tool would be useful.
The algorithm we proposed, on the other hand, has a clear interpretation with solid mathematical supports. {Specifically, the scattering transform we choose has a solid harmonic analysis support, and it is motivated by capturing the essence of CNN. In this sense, our approach combines the ``neural network'' and diffusion geometry to design features for the sleep dynamics. Note that besides knowing that the scattering transform allows us capturing the dynamical information from the spectral domain in a nonlinear way, we cannot claim that we have a full interpretation of what physiological information the scattering transform tells us. This is a physiological problem we need to explore in the future work.}
Moreover, a peculiar property of medical databases, the ``uncertainty'', deserves more discussion. 
Take the sleep dynamics studied in this paper as an example. It is well known that the inter-expert agreement rate is only about 80\% for normal subjects, not to say for subjects with sleep problems \cite{Norman2000}. With this uncertainty, a supervised learning algorithm {\em might} learn both good and bad labels. On the other hand, the {proposed feature extraction} approach is independent of the provided labels, and the chosen {scattering} features all come from the EEG signal, and speak solely for the sleep dynamics but not the labels. To some extent, the ``uncertainty'' issue is less critical via {our feature extraction} approach, since the uncertain labels are not taken into account in the feature extraction step. 

Since both supervised and unsupervised approaches have their own merits, it is natural to seek for a way to combine both. We are exploring the possibility of combining DNN and the proposed feature extraction techniques, and the result will be reported in the future work.

\subsection{Limitation and future work}
Despite the strength of the proposed method, the discussion is not complete without mentioning its limitations.
While we {test} the algorithm on two databases, and compare our results with those of state-of-the-art algorithms, those databases are small. To draw a conclusion and confirm its clinical applicability, a large scale and prospective study is needed.

Although extended theoretical understandings of applied algorithms, {including the scattering transform and diffusion-based sensor fusion} have been established in the past decade, there are still open problems we need to explore.
For example, there are open questions on the the statistical property of scattering transform on EEG. While alternating diffusion map and multiview DM are designed for the same ``sensor fusion '' purpose, they are developed under different motivations, and the consequence is never discussed. The relationship between alternating DM and multiview DM {when there are more than 2 channels is also not clear at this moment. It is well known that DM leads to an embedding free up to rotation, so aggregating intrinsic sleep features from different subjects is not a suitable approach. While the proposed approach, constructing a universal coordinate by applying DM on the pool of all scattering EEG spectral features from all subjects, leads to a satisfactory result, in general we run into the computational issue. Finding a systematic numerical solution to this problem is out of the scope of this work. The imbalanced dataset is a general challenge in data science. May we take more physiological knowledge to conquer this limitation is an interesting direction to explore.}
Another open problem is how to further take the temporal information, like the 3-minute rule, into account when we deal with the inter-individual prediction. The above mentioned limitations will be studied and reported in the future work.

\section{Conclusion}

A novel sleep stage prediction algorithm with an accurate automatic sleep stage annotation is proposed. 
The main novelty is organizing {scattering EEG spectral} features by the unsupervised diffusion geometry based sensor fusion framework, which has a solid mathematical backup. 
In addition to providing a visualization of the sleep dynamics, the prediction result is compatible with several state-of-the-art algorithms based on neural network. 
In addition to visually seeing the relationship between different sleep stages, the prediction result suggests its potential in practical applications.  
The flexibility of the proposed sensor fusion framework allows researchers to further explore other research problems that have complicated dynamics. For the reproducibility purpose and its application to other problems, the Matlab code will be provided via request.

\section{Funding}
G.-R. Liu is supported by Ministry of Science and Technology (MOST) grant \texttt{MOST 106-2115-M-006 -016 -MY2}.
Y.-L. Lo is supported by MOST grant \texttt{MOST 101-2220-E-182A-001, 102-2220-E-182A-001, 103-2220-E-182A-001 and MOST 104-220-E-182-002}.
Y.-C. Sheu is supported by MOST grant \texttt{MOST 106-2115-M-009-006} and NCTS, Taiwan.

\bibliographystyle{elsarticle-num}
\bibliography{reference}

\end{document}